\documentstyle[12pt,doublespace]{article}
\newcommand{\drawsquare}[2]{\hbox{%
\rule{#2pt}{#1pt}\hskip-#2pt
\rule{#1pt}{#2pt}\hskip-#1pt
\rule[#1pt]{#1pt}{#2pt}}\rule[#1pt]{#2pt}{#2pt}\hskip-#2pt
\rule{#2pt}{#1pt}}
\newcommand{\PSbox}[3]{\mbox{\rule{0in}{#3}\includegraphics{#1}\hspace{#2}}}
\newcommand{\Yfund}{\raisebox{-.5pt}{\drawsquare{6.5}{0.4}}}
\newcommand{\Yasymm}{\raisebox{-3.5pt}{\drawsquare{6.5}{0.4}}\hskip-6.9pt%
        \raisebox{3pt}{\drawsquare{6.5}{0.4}}}
\newcommand{\Ysymm}{\raisebox{-.5pt}{\drawsquare{6.5}{0.4}}\hskip-0.4pt%
        \raisebox{-.5pt}{\drawsquare{6.5}{0.4}}}
\newcommand{\beq}{\begin{equation}}
\newcommand{\eeq}{\end{equation}}

\font\zfont = cmss10 
\newcommand\ZZ{\hbox{\zfont Z\kern-.4emZ}}
\def\inbar{\vrule height1.5ex width.4pt depth0pt}
\def\IC{\relax\hbox{\kern.25em$\inbar\kern-.3em{\rm C}$}}

\begin{document}

\begin{titlepage}

\begin{center}

\hfill  MIT-CTP-2826

\hfill hep-th/9902118

\bigskip

\bigskip
\vspace{3\baselineskip}

{\Large \bf Marginal Deformations from Branes}

\bigskip

\bigskip

{\bf Joshua Erlich, Amihay Hanany and Asad Naqvi}\\

\bigskip
{ \small \it Center for Theoretical Physics,
		
Massachusetts Institute of Technology, Cambridge, MA 02139, USA }

\bigskip

{\tt jerlich@ctp.mit.edu, hanany@ctp.mit.edu, naqvi@ctp.mit.edu}

\bigskip

\vspace*{1cm}
{\bf Abstract}\\

\end{center}

\noindent
We study brane configurations for four dimensional 
${\cal N}$=1 supersymmetric gauge theories with quartic superpotentials which
flow in the infrared to manifolds of interacting superconformal fixed points. 
We enumerate finite ${\cal N}$=2 theories, from which a large class of
marginal ${\cal N}$=1 theories descend.
We give the brane descriptions of these theories in Type IIA and Type IIB
string theory.  The Type IIB descriptions are in terms of D3 branes in 
orientifold and
generalized conifold backgrounds.  We calculate the Weyl and Euler
anomalies in these theories, and find that they are equal in elliptic
models and unequal in 
a large class of finite ${\cal N}$=2 and marginal ${\cal N}$=1 non-elliptic
theories.
\bigskip

\end{titlepage}

\section{Introduction} 

Finite theories in four dimensions have a rich structure, much of
which has been a subject of recent interest.
Finite gauge theories having vanishing $\beta$ functions and anomalous
dimensions are conformal, and without divergences when
a perturbative expansion is valid.  Such theories contain
dimensionless parameters which are independent of scale  
and couple to marginal operators.
Finite theories often belong to a continuous manifold of scale invariant
theories.  The flow along these theories is characterized by a number of
marginal operators equal to the dimension of the manifold of fixed points.
Those marginal operators which when added to the action of a conformal theory
generate a flow along the manifold of fixed points are called exactly marginal
operators \cite{LS}.
Certain four dimensional field theories, although having a 
manifold of conformal fixed points, do not
have the property that all anomalous dimensions are zero.
These theories are not
finite by the definition above, but in the infrared they contain some number 
of exactly marginal operators.

One way of generating theories with marginal deformations is to start with a 
theory with negative $\beta$ functions, and
perturb it by an irrelevant operator in the ultraviolet.
In some cases, upon flowing to the infrared the irrelevant deformation develops
a negative anomalous dimension in such a way that it becomes marginal.

We study ${\cal N}$=1 supersymmetric gauge theories with quartic
superpotentials and marginal deformations in the infrared.  A large class of
such theories are
obtained from finite ${\cal N}$=2 theories by integrating out the adjoint 
chiral multiplet \cite{LS,karch}.  We give brane descriptions
for these theories and identify the exactly marginal operators with motions of 
NS5 branes in Type IIA configurations, or with NS two-form
fields in the Type IIB picture.  We
enumerate all finite ${\cal N}$=2 theories with one and two factors of
classical gauge groups, and discuss generalizations.  

In Section~\ref{sec:Q^4} we discuss, from a field theoretic point
of view, the conditions for existence of exactly marginal operators.  
In Section~\ref{sec:Branes} we review configurations of intersecting
NS5 branes and D4 branes in the 
presence of D6 branes and O6 plane backgrounds
in Type IIA string theory. We suggest a simple criterion which 
when imposed on the brane configurations, gives field
theories with exactly marginal operators.
For a certain class of brane configurations (the elliptic
models) in Type IIA string theory, we give a T-dual picture
in Type IIB. 
We generalize the results of \cite{KW,uranga,DM} to more
complicated conical singularities, and discuss orientifolds in ${\cal
N}$=1 and ${\cal N}=2$ theories, generalizing the analysis in \cite{PU}.  In
Section
~\ref{sec:conformal} we list all conformal ${\cal N}=2$ theories with
one or
two classical gauge groups and give the corresponding brane descriptions.
In Section~\ref{sec:sugra} we present a more general analysis of 
supergravity descriptions, and argue that a large class of
non-elliptic models which have a Type IIA description
can not satisfy the supergravity condition relating the Euler and Weyl 
anomalies of theories with supergravity descriptions \cite{gubser}, while all 
elliptic models considered are shown to satisfy that relation.

\section{${\cal N}=1$ theories with $Q^4$ type superpotentials}
\label{sec:Q^4}
Following the discussion of Leigh and Strassler \cite{LS} we argue that
four-dimensional
${\cal N}$=2 supersymmetric theories with vanishing one loop $\beta$ function
have associated with them ${\cal N}=1$ supersymmetric 
theories which have a manifold
of fixed points.  These ${\cal N}$=1 theories are obtained from the 
${\cal N}$=2 finite theories by integrating out the adjoint chiral multiplet,
giving rise to a quartic superpotential in the low energy theory.  The theory
with quartic superpotential has a line of fixed points:  In the space of 
coupling parameters of the theory, there is a line along which
the theory is conformally invariant. At
each fixed point, the Lagrangian of the theory can be deformed in such
a way that the theory remains conformally invariant.  The operator which
can be added to the Lagrangian to deform the theory in this way is said
to be an exactly marginal operator. 

The exact NSVZ $\beta_g$ function \cite{NSVZ} for the gauge coupling in a 
supersymmetric gauge theory with chiral superfields $\phi_i$ in 
representations 
$r_i$ with Dynkin indices $T(r_i)$ and anomalous dimensions $\gamma_i$ (the
normalization for the anomolous dimensions here is a factor of two larger
than in some other references), and
adjoint Dynkin index $G$, is proportional to \begin{equation}
\beta_g\propto 3G-\sum_iT(r_i)[1-\gamma_i] \label{eq:ls1}.\end{equation}
The couplings in the superpotential, which are schematically of the form
$W=h\phi_1\phi_2\cdots\phi_k$ are renormalized according to a similar equation
for their $\beta_h$ functions given by \begin{equation}
\beta_h\propto \sum_i(d_i+\frac{1}{2}\gamma_i)-3, \label{eq:ls2}\end{equation}
where $d_i$ is the canonical dimension of the field $\phi_i$.
The theory is at a fixed point when all of the 
$\beta$ functions in the theory vanish.
The anomalous dimensions are functions of the couplings, so each equation
$\beta_g=0$ or $\beta_h=0$ provides one relation between the couplings.
Generically there will be isolated solutions to these equations, or none at
all, but if the equations for the $\beta$ functions in terms of the $\gamma_i$
are linearly dependent then there may be a manifold of fixed points of 
codimension equal to the rank
of the set of linearly independent $\beta$ functions.  In that case there are 
marginal operators along the manifold of fixed points which deform the infrared
theory from one fixed point to another.

For ${\cal N}=2$ theories with massless hypermultiplets $Q_I$ in 
representations $r_I$ of the gauge group, vanishing of the one loop 
$\beta_g$ function is
sufficient for the theory to be exactly finite. For any value of the 
gauge coupling $g$, the theory is conformally invariant and so has 
a line of fixed points generated by the Tr($F^2$) operator. The anomalous
dimensions of all the fields vanish along this line and the theory is 
finite. 

We can now give mass to the adjoint chiral multiplet.  With ${\cal N}=1$ 
chiral multiplets $Q^I,\tilde{Q}^I$ 
and adjoint chiral
multiplet $\Phi$, the superpotential 
is of the form $W=h\sum_I\,{\rm Tr}\,Q^I\Phi \tilde{Q}^I+m{\rm Tr}\,\Phi^2$, 
where the trace is over the gauge indices, and $h$ is proportional to the gauge
coupling $g$.  
%
Integrating out the adjoint gives \begin{equation}
W=-\frac{h^2}{4m}(\sum_I\,{\rm Tr}\,Q^I\tilde{Q}^I)^2.\label{eq:W}
\end{equation}
Depending on the structure of the adjoint $\Phi$ it may be necessary to
add Lagrange multipliers to impose constraints on $\Phi$, e.g. tracelessness.
In that case 
the exact form of $W$ will change, but the $Q^4$ structure will remain
the same, which is the important aspect for our purposes.  The $Q^4$ coupling 
will flow to some value $\tilde{h}$ in the far infrared that will depend on
the dynamical scale and adjoint mass.
We can relax the condition that the contribution from each term in the
superpotential have the same coupling (assuming no flavor symmetry), 
and consider the possibility that
a fixed point with all of the couplings equal belong to a continuous family 
of conformal theories with $W=\sum_{I<J}
\,h_{IJ}({\rm Tr}\,Q^I\tilde{Q}^I)({\rm Tr}\,Q^J\tilde{Q}^J)$.
If the parent ${\cal N}=2$ theory has vanishing one loop $\beta_g$ function, 
then
the $\beta$ functions of the corresponding ${\cal N}=1$ theory are, according
to (\ref{eq:ls1}) and (\ref{eq:ls2}),\begin{eqnarray}
\beta_g&=&G+\sum_I T(r_I)\gamma_I \nonumber \\
\beta_{h_{IJ}}&=&1+\gamma_I+\gamma_J. \label{eq:beta} \end{eqnarray}
At a fixed point, $\beta_g=\beta_{h_{IJ}}=0$.  These are $n(n+1)/2+1$ linear
equations for the $n$ {\em a priori} unrelated anomalous dimensions.  The
diagonal components $I=J$ of (\ref{eq:beta}) completely determine the value
of the $n$ anomalous dimensions.
The solutions are $\gamma_I=-\frac{1}{2}$, and the vanishing of $\beta_g$
then follows from the vanishing of the one loop beta function of the
parent ${\cal N}=2$ theory.  Hence, the condition that $\beta_g$ vanish is
redundant and we find a line of fixed points.  In fact, each of the off
diagonal $\beta_{h_{IJ}}=0$ equations is linearly dependent on the diagonal
equations, so if there are $n$ fields
$Q^I$ then there is a $n(n-1)/2+1$ dimensional manifold of
fixed points.  The fixed points of the theory 
with superpotential (\ref{eq:W}) lie on a 
one dimensional submanifold.  There is a linear combination of the gauge 
coupling $W_\alpha W^\alpha$ and the superpotential $W$ which varies over
the line of fixed points and deforms the theory along it.

The simplest example of this phenomenon is obtained from ${\cal N}$=2 
SU($N$) Yang-Mills theory with $N_f$ hypermultiplets $Q^I$ in the
fundamental representation of the gauge group.  Upon integrating out the
adjoint, taking into account the tracelessness condition, the effective
superpotential is \begin{equation}
W\sim Q^I_\alpha\tilde{Q}^\alpha_JQ^J_\beta\tilde{Q}^\beta_I-
\frac{1}{N_c}Q^I_\alpha\tilde{Q}^\alpha_IQ^J_\beta\tilde{Q}^\beta_J.
\end{equation}
This theory was studied in \cite{LS} in relation to ${\cal N}$=1 duality.

In the following we review brane configurations involving intersecting
NS5 branes and D4 branes in Type IIA string theory 
for these types of models and
motivate a relation between marginal deformations and NS5 brane motion. We 
enumerate theories with a small number of gauge group factors 
which behave similarly to the theories discussed above and give their 
corresponding brane constructions in Type IIA and Type IIB string theory.  We
will also see that only a large class of these types of theories have
unequal Weyl and Euler anomalies, and following \cite{gubser,henningson}
therefore do not have useful supergravity descriptions in the sense of 
Maldacena's conjecture \cite{maldacena}.


\section{Brane configurations in Type II string theories}
\label{sec:Branes}
In this section we identify marginal operators with deformations of 
brane configurations or geometry in Type II string theories. We suggest
that translations of straight NS5 branes in Type IIA configurations
correspond in the field theory to 
motion along a manifold of fixed points in the infrared.
Much work has been done on understanding non-perturbative
effects in field theories from string theory. The three methods which have
been employed for this purpose rely on: (a) string theory in a 
non-trivial geometry; (b)
the world volume theory of branes probing a non-trivial (typically singular)
geometry; and (c)
the world volume theory of branes in a brane configuration in a flat
background. These different descriptions of the field theories can often be
related by various dualities. In the following we review configurations
of intersecting branes in Type IIA string theory in flat Minkowski space
for the theories alluded to in the previous section, and
then discuss the T-duality which maps these configurations to 
D-branes probing singular geometries in Type IIB string theory.  
\subsubsection*{Type IIA brane constructions and marginal deformations}
Configurations of intersecting D4 branes and NS5 branes in Type IIA
string theory in flat Minkowski space are well studied (for a review
see \cite{gk}). In this paper, we will be using NS5, NS$5^{'}$, 
D4 and D6 branes
in the following directions:
\newline
\begin{center}
\begin{tabular}{c|c c c c c c c c c c}
& 0&1&2&3&4&5&6&7&8&9 \\ \hline
NS5 & + &+&+&+&+&+& & & &  \\
NS$5^{'}$ & + & + & + & + & & & & & + & + \\
D4 & + & + & + & + &  &  & $ [$---$] $ &  & & \\
D6 & + & + & + & + &  &  &  & + & +&+ \\
\end{tabular}
\end{center}

\begin{figure}[h]
\centering
\PSbox{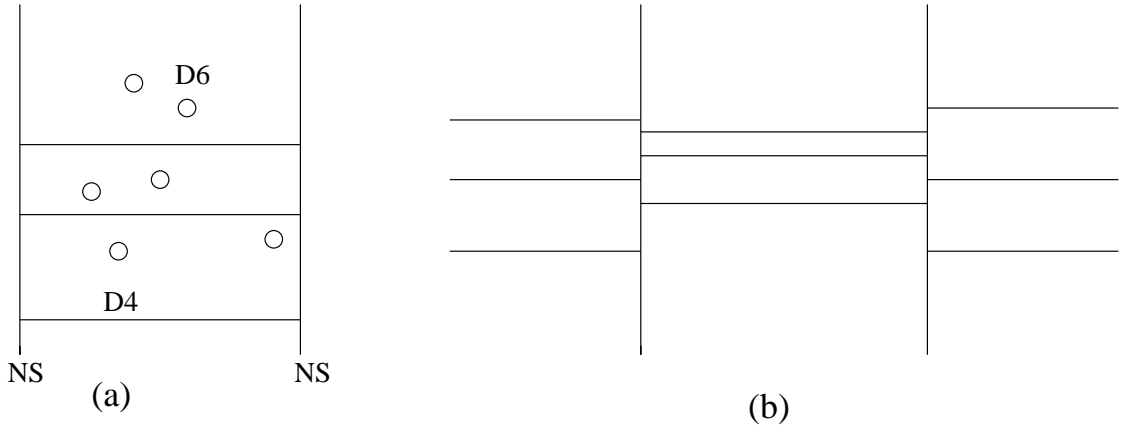}{5.5in}{2.5in}
\caption{SU($N$) with $2N$ \protect\Yfund~ hypermultiplets.}
\label{fig:SU($N$)}
\end{figure}

The configuration shown in Fig.~\ref{fig:SU($N$)}
consists of NS5, D4 and D6 branes.

The D4 branes end on NS5 branes and are of finite extent $L$ in the
$x_6$ direction. The NS5 and D4 branes are at a point in $x_7,x_8,x_9$. 
If there are $N$ D4 branes and $N_f$ D6 branes, 
the field theory on the world volume of the D4 branes at length
scales much larger than $L$ is a four dimensional ${\cal N}=2$ SU$(N)$
gauge theory with $N_f$ hypermultiplets in the fundamental representation
(corresponding to the strings 
ending on the D6 branes and D4 branes). Instead of using D6 branes, flavors
can be added with $N_f$ semi-infinite D4 branes oriented in the same 
direction as
the other D4 branes. The two ways of adding flavors to the gauge 
theory can be related by passing the D6 branes through the NS5 branes, at
which point a D4 brane is created stretching between the D6 and
the NS5 branes \cite{HW}.

As discussed in \cite{witten},
the force exerted by the D4 branes on NS5 branes causes the 
NS5 branes to bend in the $x_6$ direction. The distance between the
NS5 branes corresponds to the gauge coupling: $1/g^2=L/\lambda$ 
where $\lambda$ is the string coupling constant. The running of 
the gauge coupling is nicely reflected by the bending of the NS5 branes. 
If the beta function of the gauge theory is zero, the NS5 branes 
can be at fixed values of $x_6$ with no bending
asymptotically. This will be the case when $N_f=2N$, with $N$ semi-infinite
D4 branes from the left and $N$ from the right of each NS5 brane (if there are
no D6 branes).  In the presence of D6 branes, the requirement for there to be
no relative bending of the NS5 branes asymptotically is that the linking
number of each NS5 brane is identical:
If $L_4^i,\,R_4^i$ are the number of D4 branes attached to the left and right
of the $i$th NS5 brane and $l_6^i,\,r_6^i$ are the number of D6 branes to the 
left and right of that NS5 brane, then the linking number 
$(L_4^i-R_4^i)+\frac{1}{2}(r_6^i-l_6^i)$ must be the same for each NS5 brane.
(O6$^\pm$ planes, if present, would contribute to linking numbers as if they
were $(\pm)$ a pair of physical D6 branes.  O4 planes passing through an 
NS5 brane change sign, and contribute ($\pm 2$) units of D4 brane charge to the
linking number of the NS5 brane through which they pass.)
If a configuration has equal nonzero linking numbers for each NS5 brane, then
sixbranes can be added past the leftmost or rightmost NS5 brane such that all
linking numbers will vanish.  The additional sixbranes do not affect the field
theory on the fourbranes since they can be moved to infinity.

If one of the NS5 branes is rotated out of the $(x_4,x_5)$ plane and
into the $(x_7,x_8)$ plane, then the D4 branes can no longer slide along
the NS5 branes and the Coulomb branch of the theory is lifted.
This corresponds to giving a mass to
the adjoint proportional to tan$\,\theta$ where $\theta$ is the angle
of rotation of the NS5 branes.  (This dependence should only be trusted for
masses well below the string scale; otherwise the adjoint might decouple
at a scale higher than the string scale, but we are ignoring all string
states.)  The field theory living on the
world-volume of the D4 branes can be obtained by integrating out the
adjoint. We get a ${\cal N}$=1 theory with a quartic superpotential.  
If we start from a theory which has
no asymptotic bending of the NS5 branes, {\it i.e.} $N_f=2N$, it
is easy to see that there is an exactly marginal operator which
generates a manifold of fixed points (as discussed in detail in 
the previous section). 

\begin{figure}
\centering
\PSbox{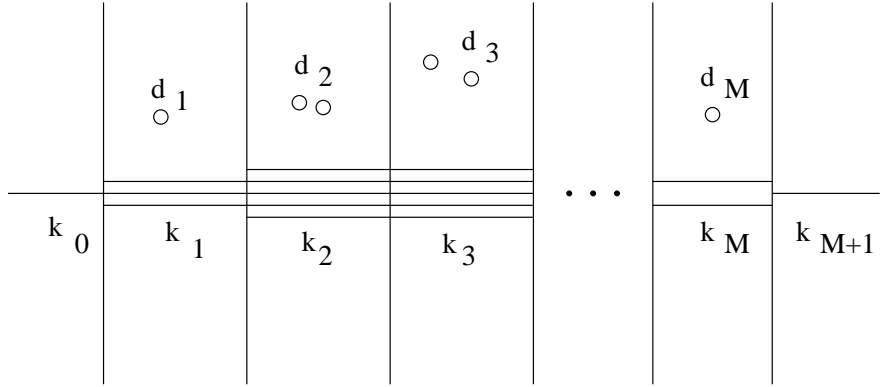}{5in}{2in}
\caption{The brane configuration corresponding to the SU$(k_1)\times$SU$(k_2)
\times\cdots\times$SU$(k_M)$ theory with bifundamentals and $k_0$ flavors
of SU$(k_1)$ and $k_{M+1}$ flavors of SU$(k_M)$.  The vertical lines
represent NS5 branes, the horizontal lines are D4 branes and the circles are
D6 branes orthogonal to the $(x_4,x_6)$ plane drawn here..
\label{fig:branes}}
\end{figure}

An obvious extension of the brane configuration discussed above is shown in 
Fig.~\ref{fig:branes}.
There are $M+1$ NS5 branes labeled
by $\alpha=1,\dots,M+1$ with $k_\alpha$ D4 branes stretched between the 
$\alpha$th and $(\alpha+1)$th
NS5 branes, and $d_\alpha$ D6 branes at points between the $\alpha$th
and $(\alpha+1)$th NS5 brane.
The gauge group of the four-dimensional theory is 
$\prod_{\alpha =1} ^M {\rm SU}(k_ \alpha )$. The matter content contains
 $d_\alpha$ hypermultiplets in the fundamental representation
of SU$(k_\alpha)$ (except for SU$(k_1)$ and SU$(k_M)$ which have
$d_1+k_0$ and $d_M+k_{M+1}$ such hypermultiplets) and bifundamental 
hypermultiplets transforming as
 ($k_1$,$\overline{k}_{2}$)$\oplus$ ($k_2$,$\overline{k}_{3}$)$\oplus$\dots 
($k_{M-1}$,$\overline{k}_{M}$). The beta function for SU$(k_\alpha)$ is
\beq
b_{0,\alpha}=-(2 k_\alpha-k_{\alpha -1}-k_{\alpha+1}-d_{\alpha}).
\label{beta}
\eeq

Rotating an NS5 brane into the $(x_7,x_8)$ plane gives a mass to the adjoint
chiral multiplets of the gauge groups to the left and right of that NS5 brane.
The contribution to the masses of the adjoints is of opposite sign to the 
left and
right of the rotated NS5 brane.  This breaks ${\cal N}$=2 to ${\cal N}$=1 and 
gives rise to a quartic
superpotential  for those fields which
transform under the gauge groups to the left and right of the NS5 brane
as discussed in Section~\ref{sec:Q^4}.

The existence of a marginal deformation in the infrared is motivated by the
absence of logarithmic bending of the NS5 branes in configurations with 
vanishing linking numbers, and hence the absence of
a scale in those theories.  The independent translations of the NS5 branes 
are intuitively expected to correspond to marginal deformations
in the infrared precisely because they are deformations not associated with
a dimensionful parameter.  Rotations of NS5 branes induce masses for the 
adjoint fields, and
are thus relevant deformations in the ultraviolet. These relevant deformations
also induce a flow along the manifold of fixed points in the infrared theory,
but do not correspond to exactly marginal operators.  More precisely, 
there is a linear combination of the gauge coupling, 
$W^\alpha W_\alpha$,
and the couplings in the infrared superpotential, $W$, which when added to 
the action of
the infrared theory deforms the theory along the line of fixed points.  At 
each fixed point there are independent marginal operators corresponding to
the gauge coupling and the terms in the superpotential (corresponding to 
motions of the NS5 branes), but by adding only one 
combination of marginal operators to the action does the theory remain 
conformally invariant.  

The proposed intuition relating deformations of brane configurations 
with straight branes and exactly marginal operators
is generally absent in a curved spacetime background because of the
scale induced by the curvature.  For example, in the presence of D6 branes and 
orientifolds, the intuition relating NS5 brane motion to marginal
operators is more tenuous, and we identify marginal deformations
only if the linking numbers of all NS5 branes, including the contributions due
to the orientifold charges, vanish.  In the ${\cal N}$=2 case, translations
of NS5 branes correspond to variation of the gauge coupling, which is indeed
an exactly marginal deformation for the finite theories.  In the 
${\cal N}$=1 case, with rotated NS5
branes, the relation between translations of NS5 branes and exactly 
marginal operators is not as immediately evident, but we will see that 
theories with
straight rotated branes generally have marginal deformations as well.
The identification of the exactly marginal operator corresponding to
the NS5 brane motion can be made at weak coupling (where it
corresponds to changing the gauge coupling---the exactly marginal
operator is Tr$F^2$). At strong coupling, the operator
corresponding to the NS5 brane motion cannot be identified easily. 
However, the existence of such an operator can be seen from the brane
picture. 

\subsubsection*{Elliptic Models}
An interesting set of models is obtained if the $x_6$ direction is
compactified on a circle of length $L$, with D6 branes between any pair of
NS5 branes.  Then $k_0=k_{M+1}$ by
definition, and the
gauge group is $\prod_{\alpha=0}^M {\rm SU}(k_\alpha) \times {\rm U}(1)$. The
extra U(1) is from uniform translation of the fourbranes along the ($x_4,x_5$)
directions.  The
hypermultiplet spectrum is as before except that the $k_0$ and $k_{M+1}$ 
fundamental hypermultiplets before compactification combine to give a
bifundamental transforming
under SU$(k_M)$ and SU$(k_0)$ and another bifundamental hypermultiplet 
transforming under SU($k_0$) and SU($k_1$). None of the hypermultiplets is 
charged under the extra U(1), so the U(1) factor decouples from the theory and 
we will not be interested in it.  
The beta function for each group is
still given by (\ref{beta}). It is easy to see that the only way
to get zero or negative beta functions for all groups is to set
$d_\alpha=0$ and $k_\alpha=N$ for all $\alpha$. 
This implies that all $b_{0,\alpha}=0$, and
there is no relative bending of the NS5 branes.

It is worthwhile to note that the relation between bending of NS5 branes
and the beta function must be reconsidered in elliptic models.  The linking
numbers are not well defined {\em a priori} because the notion of left and 
right is imprecise on the circle.  However, the relation can be made
more precise by introducing a fundamental domain on the circle.  If the circle
is cut at any point and treated as a theory on a non-compact background,
then the linking numbers on any NS5 brane must be equal for the theory to
be finite.  Choosing a different fundamental domain, {\em i.e.} cutting
the circle at a different location, will lead to different
linking numbers in general, but the equality of linking numbers is 
unaffected by this choice.  Furthermore, the beta functions are related to
relative differences between linking numbers of neighboring NS5 branes 
just as in the non-compact case.  On the circle, marginal deformations are
related to both translations of NS5 branes
along the circle and variations of the radius of the compact direction.

\subsubsection*{Type IIB descriptions}
For the elliptic models we can perform a T-duality 
along the compact $x_6$ direction. A configuration with a set of 
$N_5$ parallel NS5 branes in Type IIA string theory is mapped
to Type IIB on A$_{N_5-1} \times R^6$, where A$_{N_5-1}$ is a 
$\IC^2/\ZZ_{N_5}$ orbifold type ALE space.
The $N$ D4 branes which wrap the compact direction are mapped to D3 branes.
So the configuration discussed above with a compact $x_6$ direction
is T-dual to $N$ D3 branes at an A$_{N_5-1}$ singularity. The D3 branes
occupy the $(x_0,x_1,x_2,x_3)$ directions and the ALE space is in
the $(x_6,x_7,x_8,x_9)$ directions. 

Turning on two-form B-fields which have non-vanishing flux over
the vanishing two-cycles at the singularity corresponds to moving the NS5 
branes in the $x_6$ direction.   In the field theory this corresponds 
to variations in the gauge couplings \cite{Douglas-Moore}.
There is a question of the ordering of the NS5 branes which we will not 
discuss here.
The formalism
for studying D branes at orbifold singularities was developed in 
\cite{Douglas-Moore}. The world volume theory on the D3 branes is determined by
the orbifold action on spacetime and Chan-Paton factors (and can be encoded
in quiver diagrams or generalizations such as in 
\cite{Douglas-Moore,DG,HH,greene}), and is
the same as the theory that
we started with in the Type IIA picture before performing the T-duality. 

The ${\cal N}$=1 theories correspond to brane configurations with rotated 
NS5 branes.   On the Type IIB side, the resolved singular space varies over
the $(x_4,x_5,x_6,x_7,x_8,x_9)$ directions, and can be described by a
type of blowup of the orbifold singularity, as discussed in
\cite{KW,uranga,DM,vonUnge} for the case of the A$_1$ type singularity.
Alternatively, as described in \cite{nekrasov...}, the space can be thought
of as a complex deformation of the orbifold singularity, re-embedded in a
weighted projective space.
In the complex deformation approach, the algebraic form of the orbifold
singularity is changed so as to smooth out the singularity.  The orbifold
singularity is invariant under a $\IC^*$ action which is absent in the
deformed curve.  By re-embedding the deformed curve in a weighted projective
space we restore the $\IC^*$ action, and the resulting curve is the generalized
conifold. More explicitly, the A$_k$ orbifold singularity is given by a
curve of the form \begin{equation}
x^{k+1}+y^2+z^2=0. \label{eq:Ak} \end{equation}
It is invariant under a $\IC^*$ action with weights $(1, \frac{k+1}{2},
\frac{k+1}{2})$ for $(x, y, z)$. 

The multiplicity of the singularity is $k$, equal to the rank of the
corresponding A$_k$ gauge group, and the singularity can be thought of as
a bouquet of $k$ spheres shrinking to zero size.  A general deformation of the
curve must have at least $k$ parameters specifying the deformation.  
A deformation is called miniversal if it is specified by a number of parameters
equal to the multiplicity $k$ of the singularity, and if any deformation is
equivalent to that deformation.  One can generate a miniversal deformation
of a singular curve by adding all polynomials to the function 
specifying the curve, modulo polynomials times the first partial derivatives
of the non-deformed curve \cite{arnold}.  For 
example, 
the complex deformation of the curve (\ref{eq:Ak}) is of the form 
\cite{nekrasov...,lopez}
\begin{equation}
x^{k+1}+\sum_{m=1}^{k}t_m\,x^{k-m}+y^2+z^2=0. \end{equation}
This deformed curve is then projectivized by introducing a new variable 
$s$ with projective weight 1, giving the A$_k$ generalized conifold,
\begin{equation}
x^{k+1}+s^{k+1}\sum_{m=1}^{k}t_m\,\left(\frac{x}{s}\right)^{k-m}+y^2+z^2=0.
\label{eq:gen-conifold}
\end{equation}
We can replace the deformation
parameters $t_i$ by angles $\theta_i$, and rewrite (\ref{eq:gen-conifold}),
\begin{equation}
y^2+z^2=\prod_{m=1}^{k+1}(s\cos\theta_m+x\sin\theta_m).
\end{equation}
In this form, we identify the angles $\theta_i$ with the rotation angles of
the NS5 branes in the Type IIA picture.
 The masses of 
the adjoints are given by the relative angles,
$m_i\propto\tan(\theta_{i+1}-\theta_i)$, at least for small relative angles 
$\theta_{i+1}-\theta_i$, and satisfy the periodicity
condition $\sum_i m_i=0$.  For 
example, the A$_1$ curve, corresponding to a pair of NS5 branes on a circle, 
interpolates between the ${\cal N}$=2 orbifold ($\theta_1=0,\theta_2=0$) and 
the ${\cal N}$=1 conifold ($\theta_1=0,\theta_2=\pi/2$).  The generalized
conifold (\ref{eq:gen-conifold})
was also shown in \cite{nekrasov...} to correspond to the moduli space
of the Higgs branch of the corresponding ${\cal N}$=1 gauge theory by
studying solutions to the $D$- and $F$-flatness equations.

\subsubsection*{Adding Orientifolds}
We can obtain a rich class of theories by adding orientifolds to the 
Type IIA brane configurations described above. 
In fact, as we will see in the next
section, the finite ${\cal N}=2$ theories  
with simple classical groups and products of two factors, and with at 
most two-index matter,
can be obtained in the Type IIA brane picture if we use orientifold
planes. We will mostly be interested in O6 planes occupying the
($x_0,x_1,x_2,x_3,x_7,x_8,x_9$) directions (parallel to D6 branes introduced
earlier) and O4 planes parallel to the D4 branes. 

Consider a collection of NS5 branes (and
their images under orientifold reflection) at some distance from an O6 plane.
D4 branes can stretch between
pairs of NS5 branes, but only in configurations symmetric under the orientifold
reflection.  For example, the configuration with $2N$ D4 branes ($N$ physical
D4 branes and their images) stretched
between a NS5 brane and its orientifold image, as
discussed in \cite{LL}, corresponds to an ${\cal N}=2$ SO$(2N)$ or
USp$(2N)$ theory, depending on the sign of the orientifold charge.  We can add
$N_f$ D6 branes and their images to the above setup.  Then we get $2N_f$
fundamental chiral multiplets under the
gauge group, or $N_f$ hypermultiplets. If the sixbrane RR 
charge of the O6$^\pm$ plane is +4(-4) (physical charge +2,(-2)), we 
get an SO$(2N)$ (USp$(2N)$) gauge
theory with USp$(2N_f)$ (SO$(2N_f)$) flavor symmetry. As before, this way
of adding flavors is equivalent to adding semi-infinite D4 branes.   
The absence of bending of the NS5 branes requires vanishing beta function in
the gauge theory.  Hence we need $N_f=2N-2$ for the
case of O$6^{+}$ and $N_f=2N+2$ for the case of O$6^{-}$.  This can be
understood on the basis of the fourbrane charge induced on the NS5 branes
by the presence of the O6 plane.  Since the orientifold carries sixbrane 
charge, it interacts with the NS5 brane as though there were sixbranes or
``anti-sixbranes'' present.  That a fourbrane charge is induced is evident
by conservation of RR charge and the fact that pulling a sixbrane through
a NS5 brane produces a fourbrane connecting them \cite{HW}.  Then in order
to balance the force of the fourbrane charges on the NS5 brane, the number of
D4 branes attached to the left and right sides of each NS5 must differ.
In other words, the linking number of each NS5 brane should be the same. 

In the presence of an O6 plane as above,
NS5 branes not stuck to the O6 plane can be rotated out of the 
$(x_4,x_5)$ plane into the
$(x_7,x_8)$ plane.  Each NS5 brane and its image are rotated in opposite
directions because the configuration has to remain symmetric under
reflection about the orientifold plane, $(x_4,x_5,x_6)\rightarrow 
(-x_4,-x_5,-x_6)$. As a result of this rotation, the D4 branes
are fixed at the origin and cannot slide between the NS5 branes anymore. 
As usual, this corresponds in the field theory to giving a mass to the 
adjoint chiral multiplet, breaking ${\cal N}$=2 to ${\cal N}=1$
and lifting the Coulomb branch of the theory. The field theory
analysis shows that by integrating out the adjoint, at low energies, 
we will see a marginal quartic superpotential in the flavor superfields.

An interesting ${\cal N}=1$ configuration is obtained when
the angle of rotation of each NS5 brane is $\pi/2$. In that case, the two 
NS5 branes
become parallel to each other and to the O6 plane and the D4 branes
can slide off between the two NS5 branes.  This shows that
there must be a field in the field theory which is becoming massless
at this point. The additional 
field was shown in \cite{CSST,hanany-karch...} to tranform
as a symmetric tensor under the SO$(2N)$ gauge group or antisymmetric
tensor under the USp$(2N)$ gauge group; hence, at this point the adjoint
chiral multiplet is substituted by the opposite type of two index tensor.
The conjecture that NS5 brane motion corresponds to exactly
marginal deformations leads to a prediction that the theory on the D4 branes
of this brane configuration has a manifold of fixed points, but only for the
USp case.  The SO theory comes from an O6$^+$ plane, whose sixbrane charge
parallel to the NS5 branes cannot be canceled.  This theory will be 
discussed further in Section~\ref{sec:conformal}.
Even though we expect there to be marginal deformations of the USp
theory, we 
cannot conclude that the manifold of fixed points of this theory is related
to the manifold of fixed points of the theory without the symmetric tensor.  
The interchange of the adjoint and
tensor fields involves passing through mass scales for these fields 
larger than the string scale.  In order to trace what happens to these theories
we would have to include the effects of string states which we have been
ignoring.  Furthermore, since rotations of NS5 branes correspond
to relevant deformations, we should not expect in general to remain
on the same manifold of fixed points after such perturbations of the theory.
Translations of NS5 branes are expected to yield exactly marginal
deformations of the theory.  It is sometimes the case that relevant 
deformations,
such as those induced by rotations of the NS5 branes, lead to a flow along
the manifold of fixed points in the infrared, but this is not generically
the case.

Next we will consider the elliptic models with O6 planes. The 
$x_6$ direction is compact: $x_6\approx x_6+2L$. In that case, 
there are two orientifold fixed planes located at the 
two fixed points of the action $x_6\rightarrow -x_6$ ($x_6=0,L$). 
The O6$^\pm$ planes carry $\pm 4$ units of sixbrane charge ($\pm 2$ physical 
units). 
As was observed in \cite{PU}, vanishing total Ramond-Ramond sixbrane charge
in Type IIA ${\cal N}$=2 brane 
configurations is necessary for finiteness of the resulting gauge theories.
This is clear given the relation between NS5 brane bending and the
$\beta$ function, and is equivalent to the condition for tadpole cancelation
in the Type IIB picture of the same theories obtained after performing
a T-duality in the $x_6$ direction.

For example, consider a ${\cal N}=2$ 
theory with two NS5 branes away from the orientifolds.  If the two O6 planes
have opposite charges, as in Fig.~\ref{sosp}, then no D6 branes
are required to cancel the sixbrane charge, and the finite theory on the D4 
branes is a SO$(2N) \times\,$USp$(2N-2)$ gauge 
theory with two bifundamental half hypermultiplets.

We can now perform a T-duality along the 
$x_6$ direction, as in \cite{PU}. The D4 branes map to D3 branes.
The two NS5 branes give a $A_1$ ALE
space in the ($x_6,x_7,x_8,x_9$) directions which corresponds
to a $\ZZ_2$ singularity at $x_6=x_7=x_8=x_9=0$. The orbifold group is
$\{1,R_{6789}\}$ where $R_{6789}$ is the reflection 
\[
R_{6789}: (x_6,x_7,x_8,x_9) \rightarrow (-x_6,-x_7,-x_8,-x_9).
\]
The T-dual of the pair of O6 planes is an O7 plane. The orientifold group is
$\{1,\Omega R_{45}\}$, where $\Omega$ is a world-sheet parity reversal, and
the action on the Chan-Paton factors is such as to produce the correct gauge
group. 
However, for this to be consistent
with the orbifold action, the correct orientifold projection should
be $\{1,R_{6789},\Omega R_{45},\Omega R_{456789}\}$. Roughly speaking, 
this corresponds to an O7 plane in the ($x_0,x_1,x_2,x_3,x_6,x_7,x_8,x_9$) 
directions and an O3 plane in the ($x_0,x_1,x_2,x_3$) directions, but these
orientifolds do not carry Ramond-Ramond charge.
The action of the orientifold/orbifold system on the
Chan-Paton factors can be determined by tadpole cancelation
\cite{PU},
which leads to the SO(2$N$) $\times$ USp(2$N-2$) gauge theory as before the
T-duality.  

We can now rotate the two NS5 branes in opposite directions.  This corresponds
to adding a mass term of the form 
$m($Tr$\Phi_1^2-$Tr$\Phi_2^2)$. Integrating out the adjoint gives
a marginal quartic superpotential for the rest of the matter. 
The Type IIB configurations corresponding to Type IIA configurations
with rotated NS5 branes
are obtained by a deformation of the ${\cal N}$=2 configuration of the type 
described above and in \cite{nekrasov...}, with the
additional complication that orientifolds are present in these theories.
We do not give more explicit constructions for this case here.  Supergravity
descriptions for D-branes at combined orbifold and orientifold singularities 
were studied in \cite{kakushadze}.

\begin{figure}[h]
\centering
\PSbox{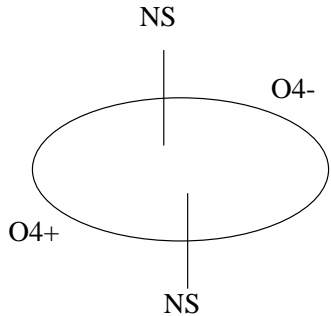}{2in}{2in}
\caption{Brane configuration with O4 plane wrapping the compact direction.}
\label{fig:O4}
\end{figure}

It is interesting now to consider the configuration with D4 branes parallel
to O4 planes  (Fig.~\ref{fig:O4}).
It is well known that the RR charge of an O4 plane changes sign as
it passes through an NS5 brane \cite{EJS,BK}. The theory on the world-volume
of the D4 brane is an ${\cal N}=2$ SO($2N$) $\times$ USp($2N$-2)  gauge
theory, which is the same theory we obtained above with O6 planes. By a 
T-duality, we apparently get D3 branes with an orientifold/orbifold 
projection which is 
the same as the one obtained for the previous case.  This is an 
interesting example of different Type IIA brane configurations which describe
the same theory on the world volume of the D4 branes, and have the same
T-dual in the compact direction.  A puzzle arises when we consider T-dualizing
the Type IIB theory along the compact direction.  It is unclear which
of the Type IIA configurations it should T-dualize to.  These types of
orientifolds are not yet understood well enough to resolve this issue, and
merit further study.

There is another configuration of two NS5 branes and O6 planes on a circle,
in which the NS5 branes are stuck in position to
the O6 planes, giving rise to an ${\cal N}$=2 SU($2N$) theory with symmetric 
and
antisymmetric tensor hypermultiplets.  The T-dual of this configuration is
again given by an O7 plane with spacetime action as above, but with a different
choice of Chan-Paton matrices, as described in \cite{PU,BK,IB,HZ}.  Note that
the configuration with the O4 plane wrapping the $x_6$ direction cannot
give rise to this theory by changing the positions of the NS5 branes.
This theory has $Q^4$ type
marginal operators as discussed earlier, but the NS5 branes in the
corresponding brane configuration (Fig.~\ref{fig:sunsymasym}) are not free to
rotate.  In this case the only marginal deformation is from changing the
circle radius. 

\section{Brane configurations for theories with exactly marginal 
operators}\label{sec:conformal}
As discussed in Section~\ref{sec:Branes}, 
construction of brane configurations
of intersecting NS5 branes and D4 branes with all the NS5 branes
having
the same linking number is a convenient way
to generate theories with exactly marginal operators.   
In this section, we study a number of such configurations.
For the 
configurations preserving 8 supercharges, which correspond to 
four dimensional ${\cal N}=2$ supersymmetric low energy theory
on the world-volume of the D4 branes, equal linking numbers of
the NS5 branes and 
the existence of an exactly marginal operator in the field theory
is implied by vanishing one-loop $\beta$ function for the field 
theory (which, in this case, implies that the theory is finite). 
The linking number criterion for brane configurations
generates almost all ${\cal N}=2$ finite configurations with
factors of classical gauge groups.  
Integrating out the adjoint chiral field from the ${\cal N}=2$ 
finite theory, we find an ${\cal N}=1$ theory with a quartic
superpotential which was shown to have an exactly marginal operator
in Section~\ref{sec:Q^4}. As we will see, some of the 
${\cal N}=1$ field theories 
with quartic superpotentials that we obtain by studying brane configurations
are not related to ${\cal N}=2$ theories by integrating out an adjoint. 
\subsection{${\cal N}=2$ finite theories with classical gauge groups}
\label{conformal1}
\subsubsection*{SU($N$) with 2$N$ \Yfund~hypermultiplets}
The brane configuration is shown in Fig.~\ref{fig:SU($N$)}. 
The linking number of each NS5 brane is zero, so the NS5 
branes are asymptotically straight. As discussed in the previous section, 
there are two equivalent ways of getting the same field theory on the 
world volume of the D4 branes -- with or without D6 branes. 
By rotating one of the NS5 branes 
in the configuration with no D6 branes, we get an ${\cal N}=1$ configuration
in which the NS5 branes still do not bend so we expect an exactly marginal
operator in the ${\cal N}=1$ field theory on the D4 branes (which is
SU($N$) gauge theory with $2N$ flavors). Rotating the
NS5 branes corresponds to integrating out the adjoint, which results in
a quartic superpotential for the field theory. The analysis in
Section~\ref{sec:Q^4}
shows that there is indeed an exactly marginal operator in the field theory.
\subsubsection*{USp(2$N$) with $(2N +2)$~\Yfund  ~ hypermultiplets}
\begin{figure}[h]
\PSbox{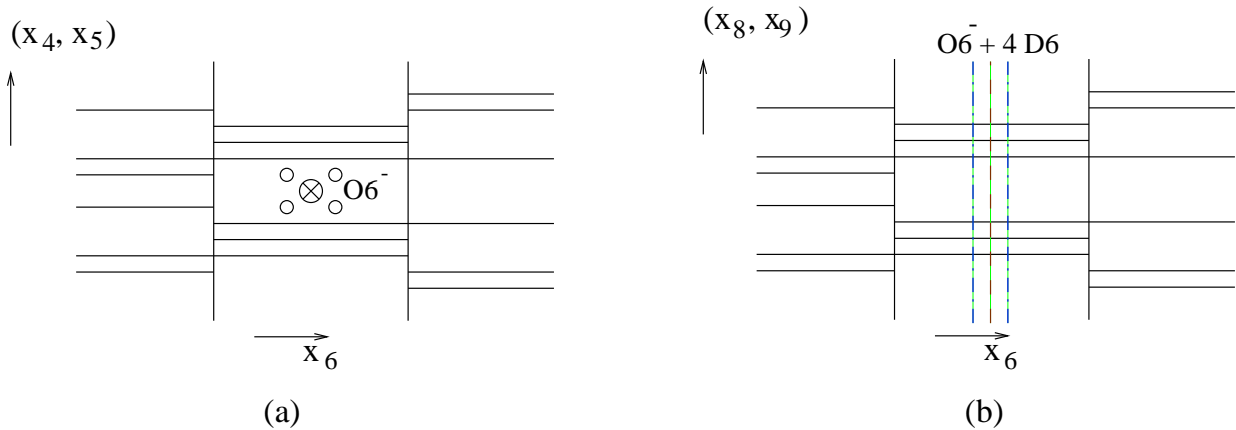}{5in}{2in}
\caption{(a) ${\cal N}=2$ USp($2N$) theory with $(2N+2)$ \protect\Yfund~ hypermultiplets. (b) ${\cal N}=1$ USp($2N$) gauge theory with 
\protect\Yasymm~ and two types of flavors, $Q$ and $f$.}
\label{b}
\end{figure}
The sixbrane RR charge of the O$6^{-}$ plane can be locally canceled
by putting 4 D6 branes on top of it. The rest of the
D6 branes can be moved in pairs, past the 
NS5 branes and to infinity to get 
a configuration with $2N$ semi-infinite D4 branes on each side 
(Fig.\ref{b}).  
The two NS5 branes can then
be rotated by an angle $\theta$ out of the $(x_4,x_5)$ plane into
the $(x_8,x_9)$ plane. This just
corresponds to integrating out the adjoint and gives a theory
with an exactly marginal operator. When $\theta=\pi/2$, an 
antisymmetric tensor chiral field, $A$, becomes massless and parameterizes
the motion of the D4 branes along the NS$5^{'}$ branes (Fig.\ref{b}). 
This theory
has two types of flavors--$2N$ ($4N$ \Yfund~ chiral multiplets) 
flavors which we call $Q$
come from the semi-infinite
D4 branes and two flavors ($4N$ \Yfund~ chiral multiplets), 
$f$ from the D6 branes. The matter
content can be summarized as:
\begin{center}
\begin{tabular}{c c c}
& USp($2N$) & \\ \hline
$A$ & \Yasymm & 1 \\
$Q$ & \Yfund & 4$N$ \\
$f$ & \Yfund & 4 \\
\end{tabular} 
\end{center}
The superpotential is
\[
W=QAQ + Q^4 + (Qf)^2 + f^4.
\]
The conditions for vanishing $\beta$ functions are:
\begin{eqnarray*}
0 & = & 3(2N+2)-(2N-2)(1-\gamma_A)-4N(1-\gamma_Q)-4(1-\gamma_f) \\
0 & = & \gamma_A+2 \gamma_Q \\
0 & = & 1 + 2 \gamma_Q \\
0 & = & 1 +  \gamma_Q + \gamma_f \\
0 & = & 1+ 2 \gamma_f. \\
\end{eqnarray*}
Only three of these equations are linearly independent implying the
possibility that the theory has a two dimensional manifold of fixed 
points and hence has two exactly marginal operators. As before, we
can identify the translations in the $x_6$ directions of the NS5 branes
with an exactly marginal operator. However, this accounts for only one 
such operator. It is not easy to check if all the quartic terms in the 
superpotential are actually present. These terms arise from
integrating out the the adjoint in the ${\cal N}=2$ theory. If
we assume that one of these terms is zero, we will get only
one exactly marginal operator from the field theory analysis
which agrees with the counting from the brane picture. However,
if all the quartic superpotential terms are non-zero and the
field theory has two exactly marginal operators,  
we might be able to see the second exactly
marginal operator by the motion of the 4 D6 branes in the $x_6$
direction. This motion is parameterized by one variable
if we require that two physical D6 branes remain on top of each other
when they move; from the field theory point of view, this means 
that the operator preserves an SO(4) 
flavor symmetry acting on the $f$'s, which 
is implied by the superpotential above. Although motion of the 
D6 branes is irrelevant
for the IR dynamics in the ${\cal N}=2$ case, here, we expect it
to be important: for example, when we move the D6 branes past the 
NS5 branes, the theory loses a flavor. So motion of the D6 brane
past the NS$5^{'}$ branes corresponds to a relevant operator
for the field theory. We propose that the motion of the D6 branes might
correspond to an exactly marginal operator when the D6 branes are between
the NS$5^{'}$ branes.
\subsubsection*{SO($N$) with  ($N-2$) \Yfund ~ hypermultiplet}
\begin{figure}[h]
\centering
\PSbox{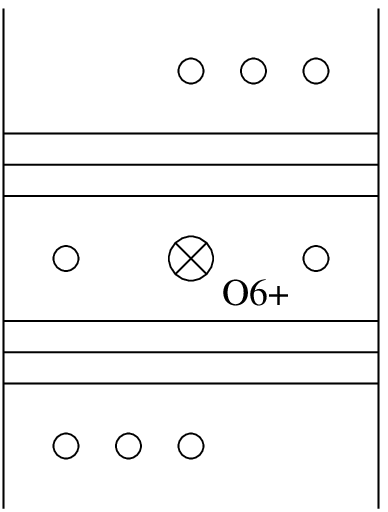}{2in}{2in}
\caption{${\cal N}=2$ SO($N$) theory with $(N-2)$ \protect\Yfund~ 
hypermultiplets.}
\label{a}
 
\end{figure}
If $N$ is odd, there is a D4 brane which cannot
move from the O6 plane. 
For the ${\cal N}=2 $ configuration (Fig.\ref{a}), the linking numbers of each NS5 brane
is zero. We can rotate the NS5 branes from
the $(x_4,x_5)$ plane into the $(x_8,x_9)$ plane by an angle 
$\theta$. The presence of the orientifold plane causes the image
of the NS5 brane to move with an angle $- \theta$. 
This corresponds to integrating out
the adjoint giving rise to a quartic superpotential. As discussed
in Section~\ref{sec:Branes}, an interesting ${\cal N}=1$
configuration arises when $\theta=\pi/2$. However, for the purpose
of generating a field theory with an exactly marginal operator, this
configuration is not very useful since the charge of the O$6^{+}$
plane cannot be canceled.
\subsubsection*{SU($N$) with $\Ysymm + (N-2)~ \Yfund $~ hypermultiplets}
\begin{figure}[h]
\centering
\PSbox{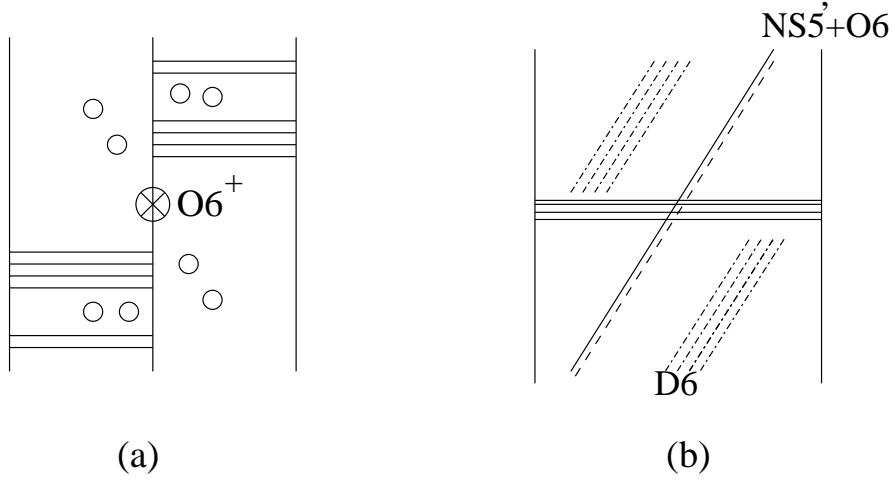}{4in}{3in}
\caption{(a)${\cal N}$=2 SU($N$) with \protect\Ysymm~ and $(N-2)$ \protect\Yfund ~hypermultiplets. (b) Replacing the center NS5 with NS$5^{'}$ to give
an ${\cal N}=1$ configuration.}
\label{c}
 
\end{figure}
The linking number of each NS5 brane
is zero which implies that the NS5 branes do not bend and the $\beta$ function
of the field theory on the D4 branes world volume is zero (Fig.\ref{c}).

An interesting chiral ${\cal N}=1$ configuration is obtained
 in which the NS5 brane on top of an O6 plane is replaced by an
NS$5^{'}$ brane in directions $(x_0,x_1,x_2,x_3,x_8,x_9)$ 
\cite{EGKT,LLL}. The NS$5^{'}$ brane, 
located at $x_7=0$ divides the
O$6$ plane into two regions --- $x_7 > 0$ and $x_7 < 0$. In such a 
configuration, the orientifold charge jumps from $-4$ to $+4$ as we cross
the NS$5^{'}$ brane \cite{EJS}. 
The part of the orientifold with negative charge
has 8 semi-infinite D6 branes embedded in it which are required by 
charge conservation \cite{EGKT}. Now $N$ D4 branes are stretched
between an NS5 brane and its image under such an orientifold (with
the NS$5^{'}$ brane and 8 semi-infinite D6 branes embedded in it)
(Fig.\ref{c}b). For
calculation of the linking number, this orientifold should act exactly
like an O$6^{+}$ plane. So we need $2(N-2)$ D6 branes ($N-2$ physical
D6 branes and their images) for the linking numbers of each NS5 branes 
to be zero. As discussed in \cite{EGKT}, the theory on the world-volume
of the D4 branes is a chiral ${\cal N}=1$ SU($N$) gauge theory with 
chiral fields:
\begin{center}
\begin{tabular}{c|c|c}
 & SU($N$)&  \\ \hline
X & \Yasymm & 1\\
$\tilde{S}$ & $ \overline{\Ysymm}$& 1 \\
 Q & \Yfund & 2$N$+4\\
 $\tilde{Q}$ & $\overline{\Yfund}$ & 2$N$$-4$\\ 
\end{tabular}
\end{center}  
It is easy to check that this theory is anomaly free -- the total anomaly
$(2N+4)-(2N-4)+(N-4)-(N+4)$ is zero. The theory has
a superpotential 
\[
W= Q \tilde{S} Q + \tilde{Q} X \tilde{Q}
\]
If we rotate the NS5 branes out of the $v=x_4+ix_5$ plane and into the
$w=x_8+i x_9$ plane by an angle $\theta$ and its image by $-\theta$,
the theory will have an adjoint $\Phi$ which will in general be 
massive except when $\theta=\pi/2$ when the adjoint becomes massless. 
The superpotential for the configuration with a generic value of 
$\theta$ is 
\[
W=Q\tilde{S}Q+ \tilde{Q} X \tilde{Q} + \Phi X \tilde{S} + \mu(\theta) \Phi^2.
\]  
For nonzero values of $\mu(\theta)$, we can integrate $\Phi$ out and 
obtain the superpotential
\[
W=Q\tilde{S}Q+ \tilde{Q} X \tilde{Q} + \frac{1}{\mu(\theta)} (X \tilde{S})^2.
\]
The equations for vanishing $\beta$ functions for the couplings of this
theory are:
\begin{eqnarray*}
0 & = & 3(2N)-(N-2)(1-\gamma_X)-(N+2)(1-\gamma_{\tilde{S}})-
(2N+4)(1-\gamma_Q)-(2N-4)(1-\gamma_{\tilde{Q}}) \\
0 & = & 2 \gamma_Q + \gamma_{\tilde{S}} \\
0 & = & 2 \gamma_{\tilde{Q}} + \gamma_X \\
0 & = & 1 + \gamma_X + \gamma_{\tilde{S}}.\\
\end{eqnarray*}
These equations are linearly dependent-- any three imply the fourth. 
So we expect an exactly marginal operator in the field theory which
is what we see from the brane picture.  
\subsubsection*{SU($N$) with $\Yasymm + (N+2)~ \Yfund$~ hypermultiplets.}
\begin{figure}[h]
\centering
\PSbox{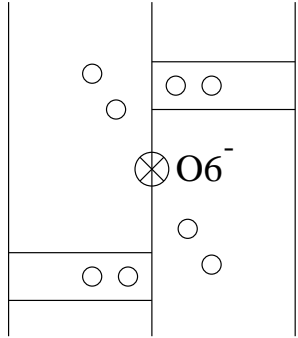}{2in}{2in}
\caption{SU($N$) with \protect\Yasymm ~ and $(N+2)$ \protect\Yfund~ 
hypermultiplets}
\label{d}
 
\end{figure}
The ${\cal N}=2$ configuration is the same as the previous model 
except that the sign of orientifold is reversed (Fig.\ref{d}). 
\subsubsection*{SU($N$) with an adjoint hypermultiplet}
\begin{figure}[h]
\centering
\PSbox{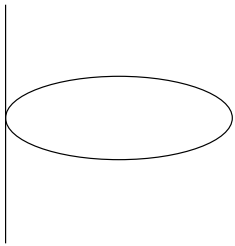}{2in}{2in}
\caption{${\cal N}=4$ SU($N$) model.}
\label{e}
 
\end{figure}
This is the ${\cal N}=4$ theory (Fig.\ref{e}). The strings 
passing through the NS5 brane with end points on the D4 branes
give rise to an adjoint hypermultiplet. There is an adjoint
chiral multiplet corresponding to the motion of the D4 branes
along the NS5 brane. So the matter content is indeed that on
an ${\cal N}=4$ theory. The superpotential couplings are also
exactly that of the ${\cal N}=4$ theory. 
Upon T-duality, we get $N$ D3 branes. 
\subsubsection*{SO($N$) with an \Yasymm~ hypermultiplet}
\begin{figure}[h]
\centering
\PSbox{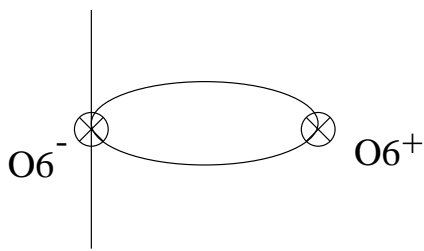}{3in}{2in}
\caption{${\cal N}=4$ SO($N$) model.}
\label{f1}
 
\end{figure}
This is the ${\cal N}=4$ theory (Fig.\ref{f1}) \cite{uranga2}.
\subsubsection*{USp($2N$) with $ \Ysymm $~ hypermultiplet}
\begin{figure}[h]
\centering
\PSbox{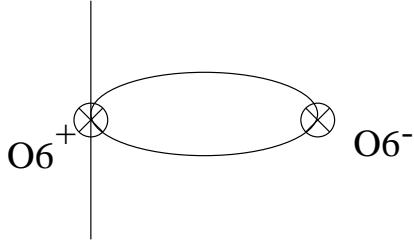}{2in}{2in}
\caption{${\cal N}=4$ USp($2N$) model.}
\label{g}
 
\end{figure}
This is an ${\cal N}=4$ theory (Fig.\ref{g}) \cite{uranga2}.

\begin{figure}[h]
\centering
\PSbox{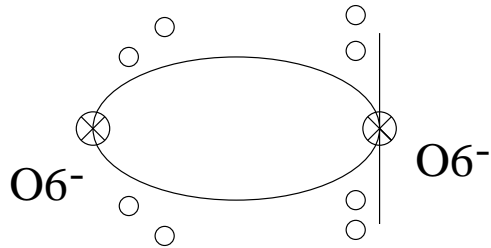}{3in}{2in}
\caption{USp(2N) with \protect\Yasymm~ and 4 \protect\Yfund ~hypermultiplets.}
\label{sp2nasymm}
  
\end{figure}

We cannot rotate the NS5 brane because of the orientifold
symmetry. However, the configuration with 
an NS$5^{'}$ brane  parallel
to the O$6^{+}$ plane preserves 4 supercharges, and we 
expect to get an ${\cal N}=1$ theory on the D4 branes
which has an exactly marginal operator. The NS$5^{'}$ brane splits 
the O6 plane into two parts and the orientifold charge jumps from $-4$
to $+4$ across the NS$5^{'}$ brane. As discussed above, we need $8$ 
semi-infinite D6 branes in the region of the orientifold with charge
$-4$. The field theory on the D4 branes has gauge group USp($2N$), and 
matter:
\begin{center}
\begin{tabular}{c c c}
& USp($2N$) & \\ \hline
$Q$ & \Yfund & 8 \\
$X$ & \Yasymm & 1 \\
$S$ & \Ysymm & 1 \\
$A$ & \Yasymm & 1 \\
\end{tabular}
\end{center}
The theory has a superpotential
\[
W=QSQ+AXS.
\]
This theory has the matter content and couplings of the ${\cal N}=2$
theory and is secretly an ${\cal N}=2$ theory. 
The one loop $\beta$ function is zero and the
theory is finite and has an exactly marginal operator. 
We can obtain this theory from a different
brane configuration shown in Fig.\ref{sp2nasymm}.
\subsubsection*{SU($N$) with $\Yasymm + \Ysymm $ ~hypermultiplets}
\begin{figure}[h]
\centering
\PSbox{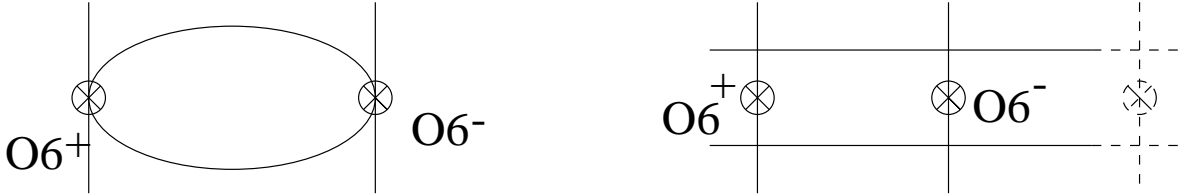}{6in}{2in}
\caption{SU($N$) with   \protect\Ysymm~   and  \protect\Yasymm   ~hypermultiplets.}
\label{fig:sunsymasym}
\end{figure}
\begin{figure}[h]
\centering
\PSbox{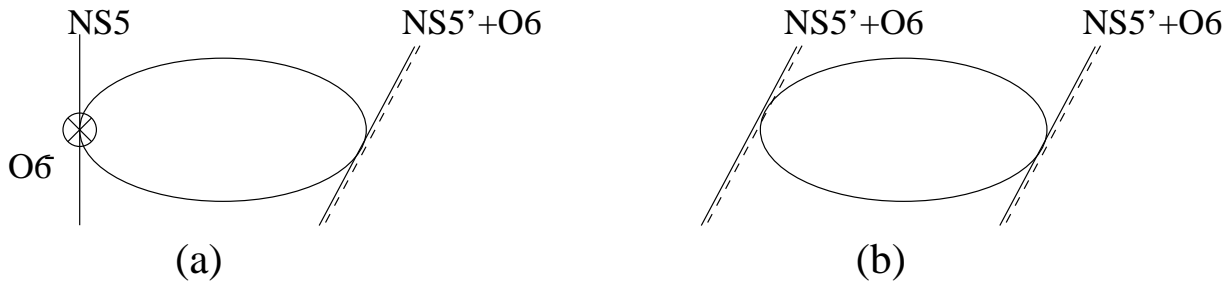}{6.5in}{2in}
\caption{${\cal N}=1$ configurations obtained by rotating NS5 branes
in Fig.\ref{fig:sunsymasym}.}
\label{sunsymasymneq1} 
\end{figure}

The $x_6$ direction is compactified on a circle. 
It is more tricky to define linking number on a circle. Finiteness implies
that the linking numbers of each NS5 branes should be the same for 
a fundamental domain of the circle. We cannot rotate the NS5 branes
continuously because of the orientifold symmetry (Fig.\ref{fig:sunsymasym}). 
However, there
are two ${\cal N}=1$ configurations we can get from this theory (shown
in Fig.\ref{sunsymasymneq1}). In Fig.\ref{sunsymasymneq1}a, the two 
NS5 branes are orthogonal to each other 
such that the NS$5^{'}$ brane (on top of the O$6^{-}$) is parallel to the O6 
planes. In fact, the NS$5^{'}$
brane splits the orientifold into two parts (corresponding to $x_7 > 0$
and $ x_7 < 0 $) and the orientifold charge jumps from $-4$ to +4 at 
$x_7=0$. For charge conservation and vanishing of the six dimensional 
anomaly, the part of the orientifold with negative charge has 
eight semi-infinite D6 branes embedded in it.  The theory on the 
D4 branes is an  ${\cal N}=1$ theory with the following chiral fields:  
\begin{center}
\begin{tabular}{c|c}
 & SU($N$) \\ \hline
$A,\tilde{A}$ & \Yasymm, $\overline{\Yasymm}$ \\
X & \Yasymm \\
$\tilde{S}$ & $ \overline{\Ysymm}$ \\
8 Q & \Yfund 
\end{tabular}
\end{center}
$A$ and $\tilde{A}$ are fields associated with the strings stretching
between the D4 branes on either side of the NS5 brane on top of the
O$6^{-}$ plane. $X$, $\tilde{S}$ and the eight $Q$'s come from the
neighborhood of the NS$5'$ brane on top of the O$6$ plane with 8 
semi-infinite D6 branes stuck to it. The theory has a superpotential
\[
W= Q\tilde{S}Q + A^4 + (X \tilde{S})^2.
\]
The $(XS)^2$ term arises by integrating out the adjoint field 
from the ${\cal N}=1$ theory shown in Fig.\ref{sunsymasymneq1}b since that
theory has a coupling of the form $X\Phi S$. Going from
configuration in Fig.\ref{sunsymasymneq1}b to that in 
Fig\ref{sunsymasymneq1}a involves giving
a mass to $\Phi$. The $A^4$ term arises by integrating
out the adjoint from the ${\cal N}=2$ theory. 
The conditions for all $\beta$ functions to vanish are:
\begin{eqnarray*}
0 & = & 3(2N)-8-3(N-2)-(N+2)+8\gamma_{Q}+2 (N-2)\gamma_{A}+(N-2)\gamma_{X}+(N+2)\gamma_{\tilde{S}} \\
0 & = & 2 \gamma_Q + \gamma_{\tilde{S}} \\
0 & = & 1+2 \gamma_A\\
0 & = & 1 + \gamma_X + \gamma_{\tilde{S}}.\\
\end{eqnarray*}
It is easy to see that these equations are linearly dependent. There 
should be one exactly marginal operator (according to the analysis in 
section~\ref{sec:Q^4}).  

\subsubsection*{SU($N$) with $2~ \Yasymm$ and $ 4 ~\Yfund$ ~hypermultiplets}
\begin{figure}[h]
\centering
\PSbox{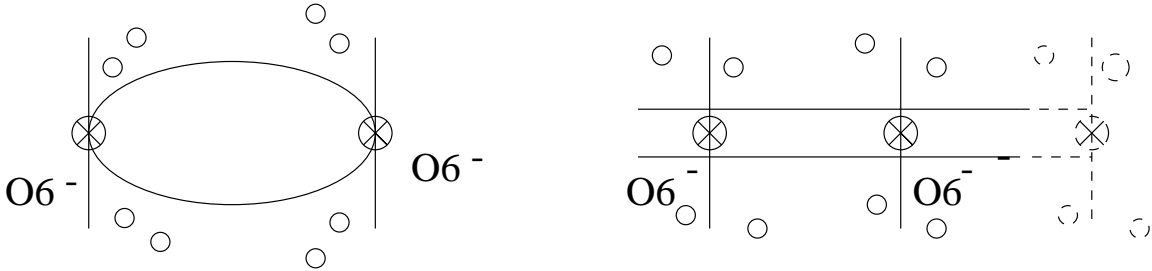}{6in}{2in}
\caption{SU($N$) with $2$~ \protect\Yasymm~ and 4 \protect\Yfund~ hypermultiplets.} 
\end{figure}
The orientifolds are O$6^{-}$ planes 
which have $-4$ units of sixbrane RR charge. Finiteness of the field 
theory implies that the RR charge in the $x_6$ direction vanishes
and that is achieved by the addition of 8 D6 branes (4 physical branes
and 4 images). These D6 branes give $4$ \Yfund~ hypermultiplets. The 
NS5 branes on each O$6^{-}$ planes gives rise to 2 \Yasymm~ hypermultiplets. 

The ${\cal N}=1$ configurations which correspond to NS5 branes being replaced
by NS$5^{'}$ are not interesting for our purposes because they will necessarily
have non-vanishing sixbrane RR charge. The orientifold charge jumps from
$-4$ to $+4$ where it intersects the NS$5^{'}$ brane. As explained above, 
we need 8 semi-infinite D6 branes embedded in the side of the orientifold
with negative charge. So the orientifold plane with 8 semi-infinite D6
branes embedded has RR sixbrane charge +4.  
\subsubsection*{USp($2N$) with  $\Yasymm$ and $4 ~\Yfund$ ~hypermultiplets}
We need four physical D6 branes to cancel the sixbrane RR charge of the
orientifolds. The T-dual is D3 branes with 4 D7 branes and an O$7^{-}$
plane. This has been discussed in \cite{sen}. 
\subsection{Product group theories with two factors of simple groups}
\label{sec:conformal2}
In this section, we construct brane configurations for product group
theories which have  manifolds of fixed
points. For simplicity, we only draw the ${\cal N}=2$ configurations.
Different ways of rotating branes in the ${\cal N}=2$
configurations in general lead to different ${\cal N}=1$ theories which
can be analyzed by the tool developed in the previous section. 
\subsubsection*{SU($N)\times $SU($M$) with matter content:}
\begin{center}
\begin{tabular}{c c c}
 SU($N$)  &  SU($M$)  & \\ \hline 
 \Yfund  & \Yfund&1  \\ 
 \Yfund  & 1&$2N-M $ \\ 
1 & $\Yfund $&$2M-N$ \\ \hline
\end{tabular} 
\end{center}
\begin{figure}[h]
\centering
\PSbox{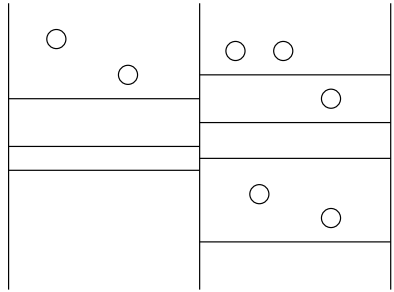}{2in}{2in}
\caption{SU($N) \times $ SU($M$) with (\protect\Yfund ~, \protect\Yfund) and flavors under each group.}
\label{susu}
  
\end{figure}
The linking number of each NS5 branes
is $(M-N)/2$ so the field theory is finite (Fig.\ref{susu}). 
Rotating one of the NS5
branes gives a ${\cal N}=1$ theory with a quartic superpotential.
Rotating the middle NS5 brane corresponds to the theory with both adjoints
integrated out, while rotating one of the outer NS5 branes corresponds to 
integrating out only one of the adjoints.
 
\subsubsection*{SU($N)\times $SU($M$) with matter hypermultiplets:}
\begin{figure}[h]
\centering
\PSbox{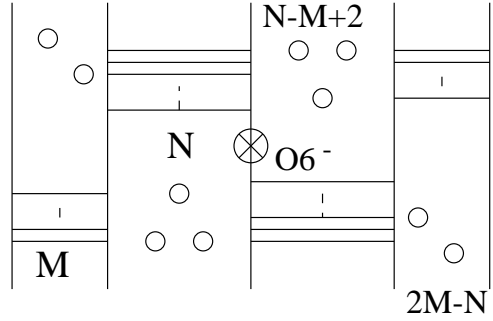}{3in}{2in}
\caption{SU$(N)\times $SU($M$) with (\protect\Yasymm~,1) and \protect\Yfund~
hypermultiplets under each group.}
\label{ab}
  
\end{figure}
\begin{center}
\begin{tabular}{c c c}
 SU($N$)  &  SU($M$) & \\ \hline 
 \Yfund  & \Yfund&1  \\ 
 \Yasymm , \Ysymm  & 1&1 \\ 
\Yfund &1&$N-M\pm2$ \\ 
1&\Yfund&$2M-N$ \\ \hline
\end{tabular} 
\end{center}
This model involves putting an NS5 brane on top of the O$6^{\pm}$ plane
(Fig.\ref{ab}).
The linking number of each NS5 brane
is $0$ precisely when the number of D6 branes and D4 branes are equal
to the numbers predicted from the vanishing $\beta$ functions for the
field theory. For O$6^{-}$, we get an \Yasymm~ under SU($N$); 
for O$6^{-}$, we get a \Ysymm~ under SU($N$).
\subsubsection*{SU($N)\times $ SU($N+2$) with hypermultiplets:}
\begin{figure}[h]
\centering
\PSbox{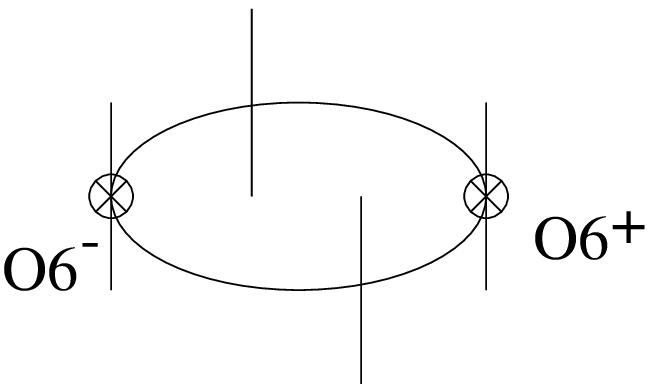}{2in}{2in}
\caption{SU($N)\times $SU($N+2$) with (\protect\Yfund, \protect\Yfund)
and ( \protect\Yasymm,1) and (1,~\protect\Ysymm)} 
\label{da}
\end{figure}

\begin{center}
\begin{tabular}{c c}
 SU($N$)  &  SU($N+2$) \\ \hline 
 \Yfund  & \Yfund \\ 
 \Yasymm  & 1 \\ 
1 & \Ysymm \\ 
\end{tabular} 
\end{center}
This is shown in Fig.\ref{da}
\subsubsection*{SU($N)\times $SU($M$) with hypermultiplets:}
\begin{center}
\begin{tabular}{c c c}
 SU($N$)  &  SU($M$)  & \\ \hline 
 \Yfund  & \Yfund&1  \\ 
 \Yasymm  & 1&1 \\ 
\Yfund &1&$N-M+2$ \\ 
1& \Yasymm  &1 \\
1& \Yfund&$M-N+2$ \\ \hline
\end{tabular} 
\end{center}
\begin{figure}[h]
\centering
\PSbox{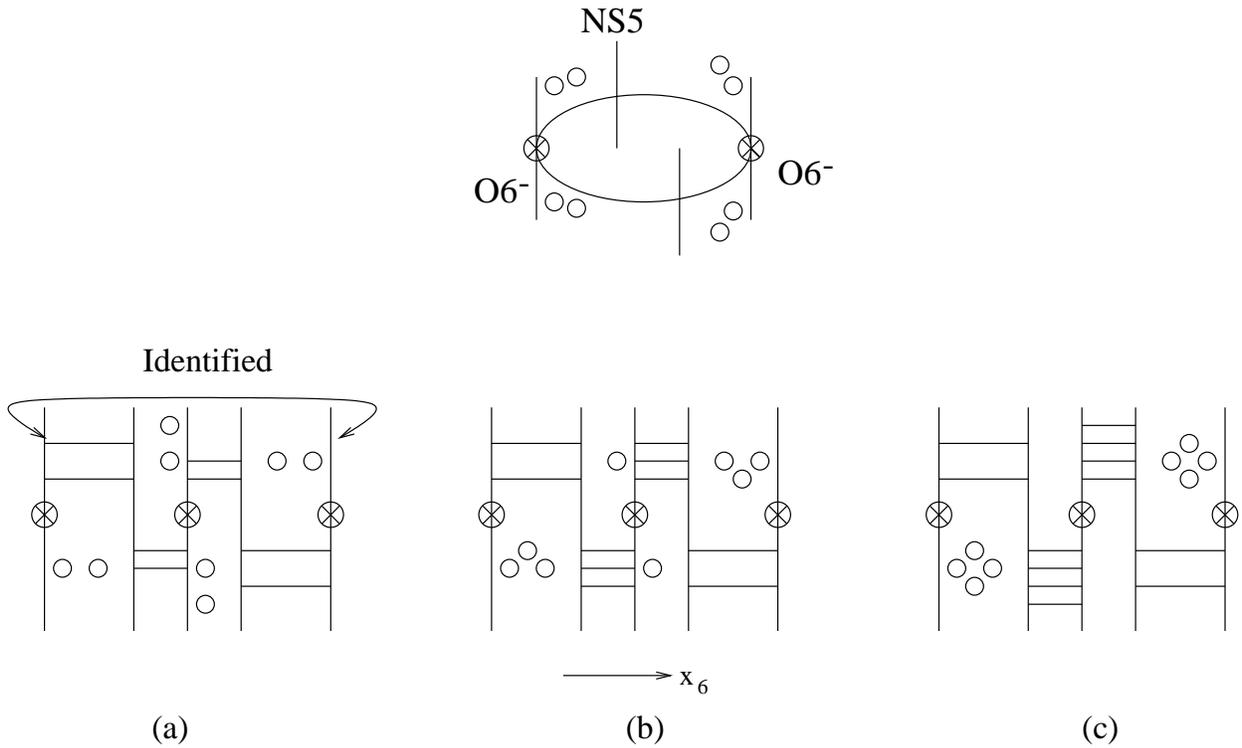}{6.5in}{4in}
\caption{SU($N) \times $ SU($M$) with (\protect\Yfund,\protect\Yfund)
and \protect\Yasymm~ and \protect\Yfund~ under each group: (a), (b) and (c) are examples of cases when $|N-M|=0,1,2$.}
\label{dada}
\end{figure}
Here, $|N-M| \leq 2 $. For each value of $|N-M|$, it can be easily
shown that the linking numbers of the NS5 branes are the same (Fig.\ref{dada}).  
\subsubsection*{SU($N)\times $SU($N$) with hypermultiplet:}
\begin{center}
\begin{tabular}{c c c}
SU($N$)  &  SU($N$)  & \\ \hline 
 \Yfund  & \Yfund&1  \\ 
 \Yfund  & \Yfund&1  \\ \hline
\end{tabular}
\end{center}
\begin{figure}[h]
\centering
\PSbox{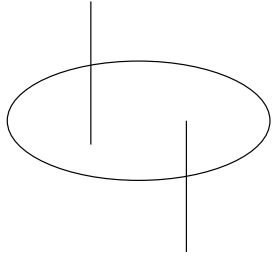}{2in}{2in}
\caption{SU($N)\times $SU($N$) with two (\protect\Yfund,\protect\Yfund) 
hypermultiplets.}
\label{baba}
  
\end{figure}
This is the elliptic model with no orientifolds and D6 branes. The 
linking number of each NS5 brane is 0 because there is an equal number
of D4 branes to the left and right of each NS5 brane (Fig.\ref{baba}). 
\subsubsection*{SO($N)\times $SU($M$) with hypermultiplets:}
\begin{figure}[h]
\centering
\PSbox{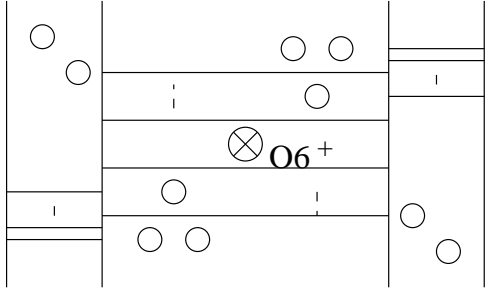}{2in}{2in}
\caption{SO($N) \times$ SU($M$) with (~\protect\Yfund~,~\protect\Yfund~) and
flavors.}
\label{cala}
  
\end{figure}

\begin{center} 
\begin{tabular}{c c c}
 SO($N$)  &  SU($M$)  & \\ \hline 
 \Yfund  & \Yfund&1  \\ 
 \Yfund  & 1&$N-M-2 $ \\ 
1 & $\Yfund $&$2M-N$ \\ \hline
\end{tabular} 
\end{center}
This is shown in Fig.\ref{cala}.
\subsubsection*{USp($2N)\times $ SU($M$)}
\begin{center}
\begin{tabular}{c c c}
 USp($2N$)  &  SU($M$)  & \\  \hline
 \Yfund  & \Yfund&1  \\ 
 \Yfund  & 1&$2N-M+2 $ \\ 
1 & $\Yfund $&$2M-2N$ \\ \hline
\end{tabular} 
\end{center}
\begin{figure}[h]
\centering
\PSbox{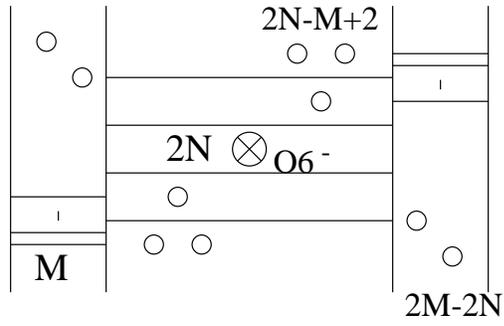}{2in}{2in}
\caption{USp($2N) \times$ SU($M$) with (~\protect\Yfund~,~\protect\Yfund~) and
flavors}
\label{calb}
\end{figure}
This is the same as
the previous case except the sign of the orientifold is reversed 
Fig.\ref{calb}.   
\subsubsection*{SO($N)\times $SU($N-2$)}
\begin{center}
\begin{tabular}{c c c}
SO($N$)  &  SU($N$-2)  & \\ \hline 
 \Yfund  & \Yfund&1  \\ 
 1  & \Yasymm&1  \\ \hline
\end{tabular}
\end{center}
\begin{figure}[h]
\centering
\PSbox{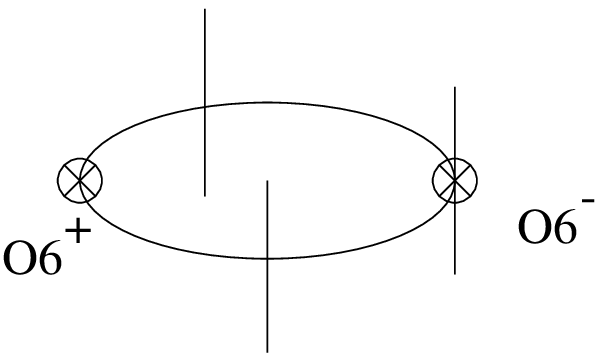}{2in}{2in}
\caption{SO($N)\times$ SU($N-2$) with (~\protect\Yfund~,~\protect\Yfund~)
and (1,~\protect\Yasymm~).}
\label{abc}
  
\end{figure}
The linking numbers of the NS5 branes are the same (Fig.\ref{abc}). 
\subsubsection*{USp($2N)\times $SU($2N+2$)}
\begin{center}
\begin{tabular}{c c c}
USp($2N$)  &  SU($2N$+2)  & \\ \hline 
 \Yfund  & \Yfund&1  \\ 
 1  & \Ysymm&1  \\ \hline
\end{tabular}
\end{center}
\begin{figure}[h]
\centering
\PSbox{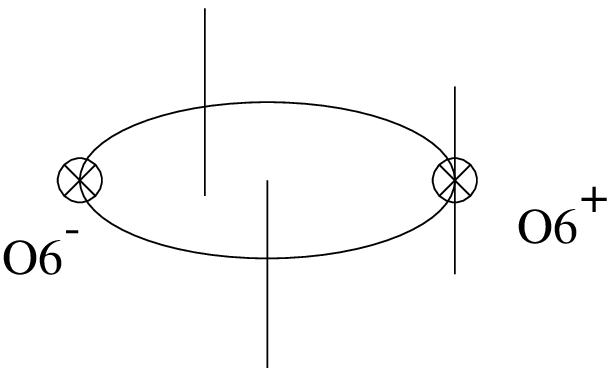}{2in}{2in}
\caption{USp($2N)\times$ SU($2N+2$) with (~\protect\Yfund~,~\protect\Yfund~)
and (1,\protect\Ysymm).}
\label{def}
\end{figure}
This is again an elliptic model --- the signs of the orientifolds
are reversed compared to the previous theory (Fig.\ref{def}). 
\subsubsection*{USp($2N)\times $ USp($2M$) $|2M-2N|\leq 2$}
\begin{figure}[h]
\centering
\PSbox{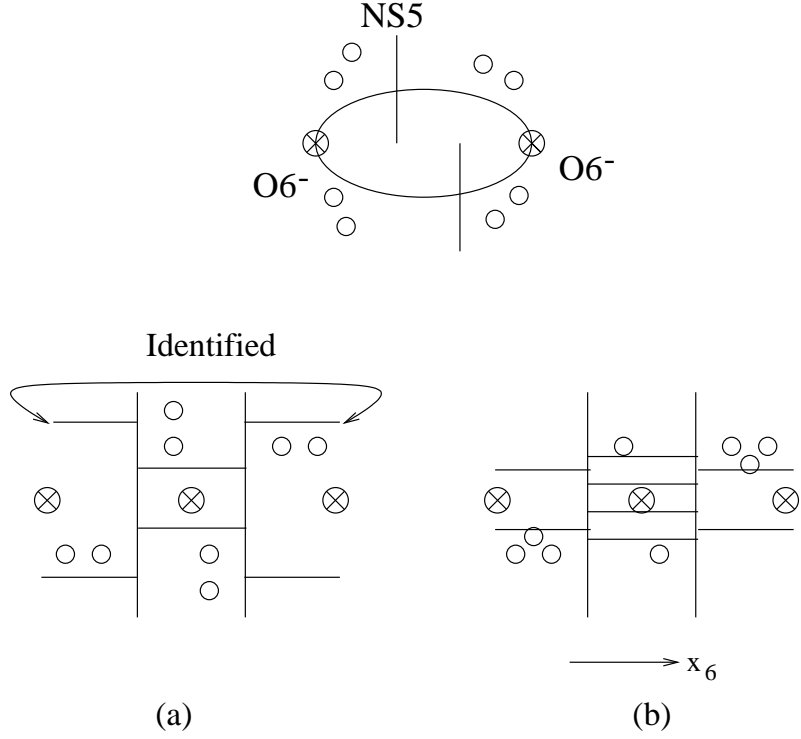}{4in}{4in}
\caption{USp(2$N)\times$USp($2M$) with (\protect\Yfund~, \protect\Yfund),
\protect\Yfund~ under each group:(a), (b) are the cases when $|2N-2M|=0,2$.}
\label{fgh}
\end{figure}
\begin{center}
\begin{tabular}{c c c}   
USp($2N$)  &  USp($2M$)  & $|2M-2N| \leq 2$\\ \hline
\Yfund  & \Yfund&1  \\ 
\Yfund  & 1&$2N-2M+2 $ \\ 
1 & $\Yfund $&$2M-2N+2$ \\ 
\end{tabular} 
\end{center}
This theory is shown in Fig.\ref{fgh}. 
\subsubsection*{USp($2N)\times $ SU($M$) $|M-2N|\leq 2$}
\begin{figure}[h]
\centering
\PSbox{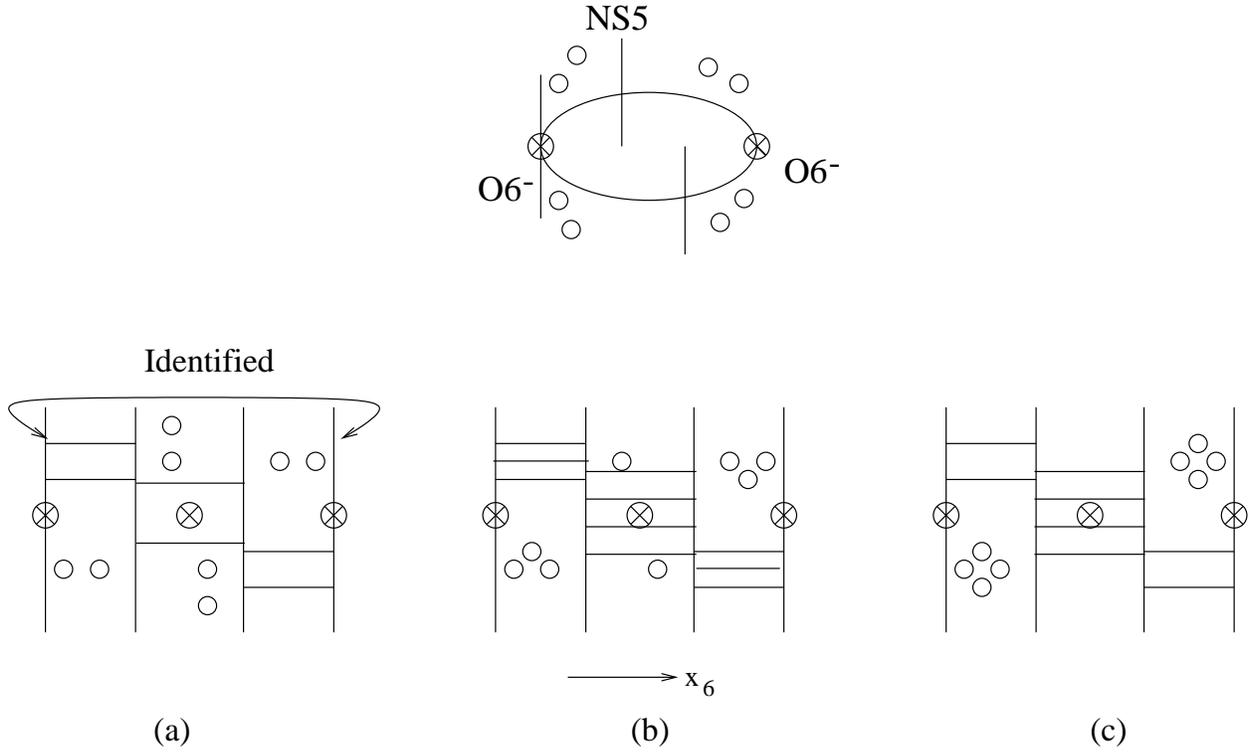}{6.5in}{4in}
\caption{USp(2$N) \times $ SU($M$) with (\protect\Yfund~, \protect\Yfund),
\protect\Yfund~ under each group and \protect\Yasymm~ under SU($M$):
(a), (b) and (c) are the cases when $|2N-M|=0,1,2$.}
\label{sht}
  
\end{figure}
\begin{center}
\begin{tabular}{c c c}   
USp($2N$)  &  SU(M)  & $|M-2N| \leq 2$\\ \hline
\Yfund  & \Yfund&1  \\ 
\Yfund  & 1&$2N-M+2 $ \\ 
1 & $\Yfund $&$M-2N+2$ \\ 
1 & \Yasymm & 1 \\ \hline
\end{tabular} 
\end{center}
This model has two O$6^{-}$ plane and has 8 D6 branes so the net
sixbrane RR charge vanishes. $|M-2N|\leq 2 $ and the different values
of $|M-2N|$ just correspond to placing the 8 D6 branes in various ways
in between the NS5 branes such that the linking numbers of each NS5 brane
still comes out the same (Fig.\ref{sht}). 
\subsubsection*{SO($N)\times $USp($N-2$) with half-hypermultiplets}
\begin{center}
\begin{tabular}{c c c}
SO($2N$)  &  USp($2N$-2)  & \\ \hline 
 \Yfund  & \Yfund&2  \\ \hline
\end{tabular}
\end{center}
\begin{figure}[h]
\centering
\PSbox{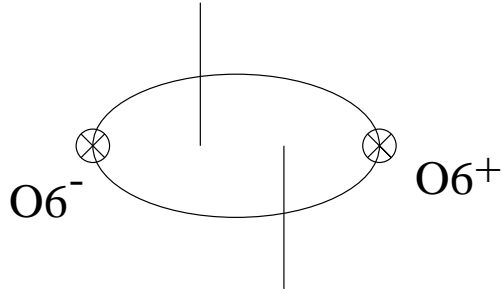}{2in}{2in}
\caption{SO($2N) \times $ USp($2N-2$) with two (\protect\Yfund~, 
\protect\Yfund) half-hypermultiplets.}
\label{sosp}
\end{figure}
The field theory has two half-hypermultiplets. The brane configuration is
in Fig.~\ref{sosp}.

\section{Supergravity Descriptions}
\label{sec:sugra}
As discussed in section \ref{sec:Branes}, we can perform a 
T-duality along the compact direction for the elliptic models in Type IIA
to get to a configuration with D3 branes in some singular geometry. 
Given the Type IIB constructions, one can 
in principle
determine a supergravity description for those theories.  This is complicated 
in
practice because of the orbifold,conifold and orientifold geometries, 
but the near
horizon geometry of the D3 branes in these backgrounds is expected to
be related to the corresponding gauge theories via the AdS/CFT correspondence
\cite{maldacena,witten2,gkp}.  It has been argued 
\cite{gubser,henningson} that the 
difference between the Weyl and Euler anomalies must vanish to leading order
in $N$ in gauge theories
which have useful supergravity descriptions.  In this section we argue that the
elliptic models admit useful supergravity descriptions, while non-elliptic 
${\cal N}$=2 finite and their descendent ${\cal N}$=1 marginal theories do not.
Consider the ${\cal N}$=2 finite elliptic model given by the brane
construction in Fig.~\ref{fig:SU(N)^M}.  The gauge theory on the four-branes
has gauge group SU$(N)^M$ with bifundamental hypermultiplets as below:

\begin{figure}[h]
\centering
\PSbox{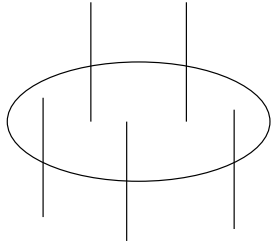}{2in}{2in}
\caption{Brane configuration for the elliptic SU$(N)^M$ theory.}
\label{fig:SU(N)^M} 
\end{figure}

\begin{center}
\begin{tabular}{c c c c c| c}
SU$(N)$  &  SU$(N)$  & SU$(N)$ & $\cdots$ & SU$(N)$ & U(1)$_R$ \\ \hline 
 \Yfund  & \Yfund & & & & 0\\ 
   & \Yfund& \Yfund & & & 0 \\
 & & $\cdots$ & & & 0\\
\Yfund & & & & \Yfund & 0\\ \hline
\end{tabular}
\end{center}
Since the fermion in the adjoint chiral multiplets should transform with the
same $R$-charge as the gauginos (by ${\cal N}$=2 supersymmetry), the adjoint 
chiral multiplets have $R$-charge 2.

At a conformal fixed point, the difference between the Weyl and Euler 
anomalies, $c-a$, is proportional to the U(1)$_R$ anomaly \cite{afgj,aefj}.  
The U(1)$_R$ anomaly is easily computed:
\begin{equation}
\partial\left<J_RTT\right> \sim \sum_i \,{\rm dim}\,R_i (r_i-1),\end{equation}
where dim$\,R_i$ is the dimension of the representation of the chiral
multiplet with $R$-charge $r_i$, and $T$ is shorthand for the stress tensor.
Since the adjoint fermions in the ${\cal N}$=2 SU$(N)^M$ elliptic model have
charge +1 and the matter fermions have charge $-1$, it is easy to see that the
condition $c-a=0$ is satisfied.  One way to see this is that the number
of adjoint fermion degrees of freedom is $2MN^2$, which is the same as the 
number of matter fermion degrees of freedom.

Now consider rotating one of the NS5 branes.  This breaks ${\cal N}$=2 to
${\cal N}$=1 with the result of assigning $R$-charge zero to the adjoint
fermions (except the gauginos, which have charge 1 by convention) and 
$R$-charge $-1/2$ to the matter multiplet fermions.  The
anomaly $c-a$ is proportional in this case to the number
of gauginos minus half the number of matter fermion degrees of freedom.  
Since now the adjoint chiral fermions (half of the ${\cal N}$=2 vector 
multiplet) do not contribute, the anomaly is
proportional to the ${\cal N}$=2 result, so $c-a=0$ in this case as well.
This way of thinking about the quantity $c-a$, as counting fermion degrees
of freedom, is useful because it can be easily generalized to the more 
complicated cases with orientifold planes.

If we add a pair of O6 planes in the ${\cal N}$=2 elliptic model with 
appropriate symmetry to accommodate the O6 reflections as described
in the previous sections, the effect is as follows:

\begin{figure}[h]
\centering
\PSbox{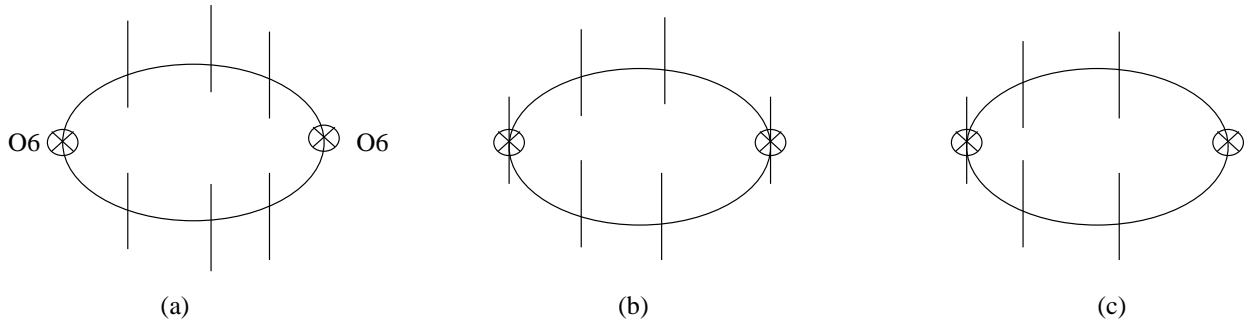}{7.5in}{2in}
\caption{Elliptic models with O6 planes.  D6 branes and the signs of the 
RR charge carried by the O6 planes is not given in the figures.  The net D6
brane charge always cancels in these theories.  Furthermore, the NS5 branes
that are not stuck to the O6 planes can be rotated symmetrically.}
\label{fig:O6} 
\end{figure}

a) If the number of NS5 branes (including the images under the O6 reflection)
is even, $N_5=2M$, and the O6 planes do not 
intersect any of the NS5 branes (Fig.~\ref{fig:O6}a), then the gauge theory 
has $M+1$ factors.
Two of the factors are SO or USp, while the rest are SU.  To order $N^2$
the adjoint of SO$(N)$ or USp$(N)$ has $N^2/2$ degrees of freedom.  Summing the
contribution from the two SO or USp factors gives $N^2$, the same as the
contribution from adjoints of the SU$(N)^M$ elliptic model.  Including the 
$M$ bifundamentals, the U(1)$_R$ anomaly again vanishes.

b)  If the number of NS5 branes is even, $N_5=2M$, and the O6 planes intersect
two of the NS5 branes (Fig.~\ref{fig:O6}b), then the gauge theory has 
$M$ SU factors, $M-1$
bifundamentals and two symmetric or antisymmetric tensors.  The symmetric and 
antisymmetric
tensors make up the difference in degrees of freedom corresponding to
the extra bifundamental in the elliptic SU$(N)^M$ theory.

c)  IF the number of NS5 branes is odd, $N_5=2M+1$, then one O6 plane 
intersects an NS5 brane and the other does not (Fig.~\ref{fig:O6}c).  
The gauge theory has $M+1$
factors, one of which is SO or USp.  There are $M$ bifundamentals and one
symmetric or antisymmetric tensor.  Since the gauginos of SO contribute half
as many degrees of freedom as those of SU, and similarly for the symmetric or 
antisymmetric tensor, the contribution of the SO or USp gauginos to $c-a$ 
cancels with the tensor, and the counting of the remaining degrees of freedom
is again like the SU$(N)^M$ theory.

Depending on the sign of the RR charges of the orientifolds, in some theories
extra D6 branes will be required to cancel the flux of sixbrane charge in the
$x_6$ direction, or equivalently for finiteness of the ${\cal N}$=2 theory.  
The D6 branes give rise to additional flavors, which are also required
in those cases for conformality.  The additional flavors do not contribute
to $c-a$ to leading order in $N$, so they were ignored in the counting above.

Alternatively, an O4 plane can wrap the $x_6$ direction, parallel to the 
D4 branes, in which case the theory is an alternating 
SO$\times$USp$\times\cdots$ theory with bifundamental half hypermultiplets.
It is easy to see that $c-a=0$ in this case, as well.

As discussed earlier, NS5 branes can be rotated in the orientifold theories 
in such a way as to
preserve the orientifold symmetry, breaking ${\cal N}$=2 to ${\cal N}$=1.
The argument regarding rotating branes in the SU$(N)^M$ elliptic models
is valid in these cases as well, and we find $c-a=0$ for the 
${\cal N}$=1 elliptic orientifold models.

For certain specific orientations of
the NS5 branes with respect to the O6 planes, additional massless degrees
of freedom appear.  For example,  as described in the last section,
if an NS5 brane and its image are parallel to the O6$^\pm$ plane, and there 
are no other NS5 branes
between them, then an additional antisymmetric (symmetric) tensor appears
for the corresponding USp (SO) gauge group,
associated with movement of the D4 branes in the $x_7,x_8$ direction 
\cite{CSST}.  The
additional tensor does not contribute to the U(1)$_R$ anomaly because by
gauge anomaly freedom the fermion in the tensor chiral multiplet has vanishing
$R$-charge.

\begin{figure}[h]
\centering
\PSbox{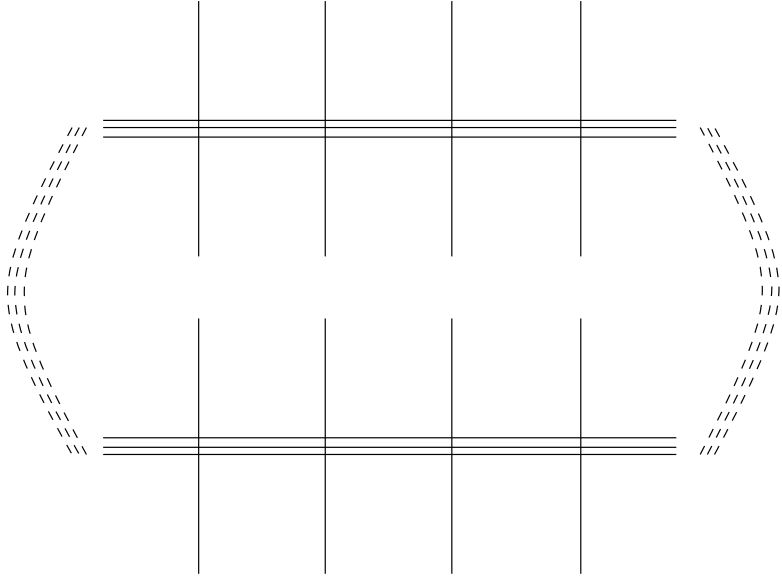}{4in}{3in}
\caption{Glueing together copies of non-elliptic theories with vanishing
net sixbrane charge on a circle.}
\label{fig:glue}  
\end{figure}

\begin{figure}[h]
\centering
\PSbox{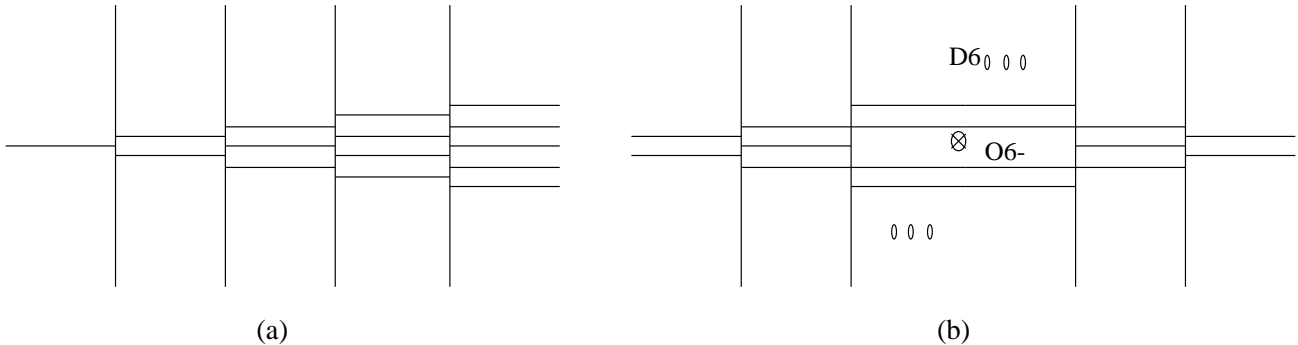}{7in}{2in}
\caption{Some non-elliptic brane configurations not included in the 
discussion of Weyl and Euler anomalies.}
\label{fig:exceptions}  
\end{figure}

The equivalence of the Weyl and Euler anomalies in the elliptic theories is
not surprising, since Type IIB configurations have been constructed which
describe them, from which one can determine their supergravity description.
However, this result can be used to prove that non-elliptic conformal or
marginal theories which have a Type IIA brane description with vanishing 
net sixbrane charge and linking numbers do not satisfy the 
condition $c-a=0$.  The argument is as follows:  Two copies of the 
Type IIA brane configuration for the non-elliptic theory can be connected
to form an elliptic model considered above if the net sixbrane charge and
NS5 linking numbers vanish
for each copy separately (Fig.~\ref{fig:glue}).  The resulting
elliptic model satisfies the condition $c-a=0$ as discussed above.  
The 
difference in the counting of degrees of freedom in this case versus
the case of the two separate copies is from additional gauginos for each of
the two new fourbrane links.  
Since the degrees of freedom of the new gauginos do not 
cancel the four additional bifundamentals, the non-elliptic theory from which
we started could not have satisfied $c-a=0$.  Theories which are not included
in this argument include those with non-vanishing linking numbers, {\em i.e.}
bending branes, theories with non-vanishing net sixbrane charge, and theories 
without semi-infinite D4 branes at both the left and right side of the 
configuration (Fig.~\ref{fig:exceptions}).

\section{A comment on brane boxes}
A similar analysis to that of the last section can be done for brane box
models.  
A large class of brane box models describes finite ${\cal N}$=4,2,1
theories \cite{HSU}.  In these theories there are bifundamentals and/or 
adjoints with
cubic superpotentials.  The ${\cal N}$=2 theories are the same as the 
SU$(N)^M$ theories described above, and the ${\cal N}$=4 theory is the usual
SU($N$) gauge theory.  The ${\cal N}$=1 models differ from ours in both
matter content and superpotential.  We briefly review their construction
and describe the restrictions on which of these theories may have a 
supergravity description in light of the results of \cite{gubser}, in 
analogy with the discussion in the previous section.

The basic brane box for four dimensional gauge theories is a  Type IIB 
brane configuration consisting of a two dimensional lattice of NS5 branes 
filled with D5 branes of finite extent in two directions.  We will consider
elliptic brane box configurations, in which the configuration is defined
on a torus.  There are two classes of such configurations: 

\begin{figure}[h]
\centering
\PSbox{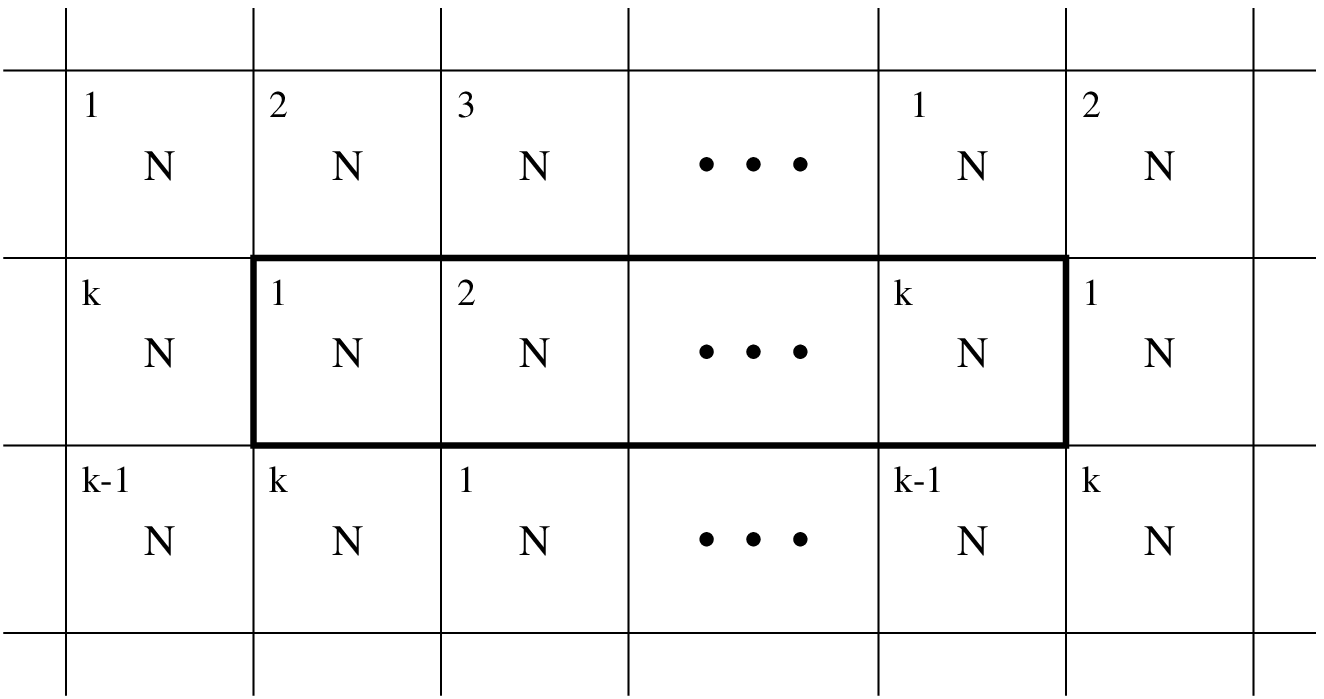}{5.5in}{3in}
\caption{Twisted $1\times k$ elliptic brane box models. There are $N$ D5
branes in each box of NS5 branes.}
\label{fig:branebox1} 
\end{figure}

a) ${\cal N}$=1 
configurations can be obtained by twisting the
torus of an ${\cal N}$=2 configuration, as in Fig.~\ref{fig:branebox1}.
There are bifundamental
chiral multiplets between pairs of neighboring and diagonally neighboring 
SU($N$) group factors labeled 1,2,$\dots$,$M$.  The grid represents the 
NS5 branes, and there are $N$ D5
branes in each box.  The matter content can be summarized as three sets of
bifundamental chiral multiplets cyclically permuted among the SU$(N)$ 
factors.  There is
a cubic superpotential consisting of gauge invariant triple products of the
bifundamentals.  

\begin{figure}[h]
\centering
\PSbox{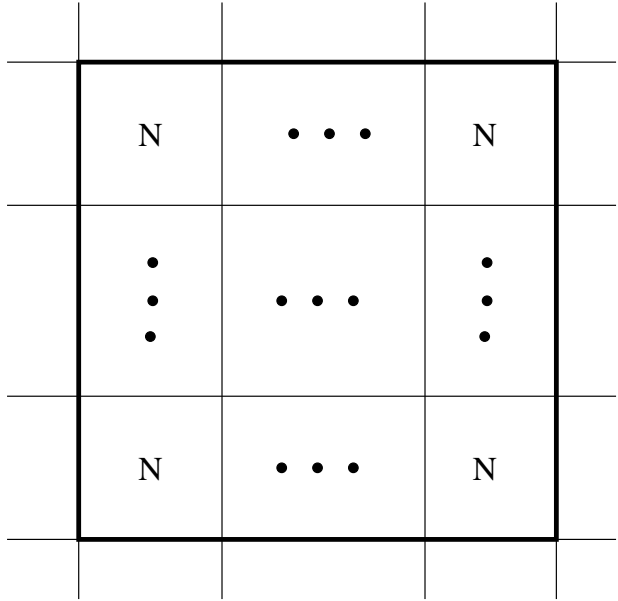}{3in}{3.2in}
\caption{Generic $k \times k^\prime$ brane box configuration.}
\label{fig:branebox2} 
\end{figure}

b)Alternatively, generic tori of $k\times k^\prime$ boxes with $k,\,k^\prime>1$
describe ${\cal N}$=1 theories, as in Fig.~\ref{fig:branebox2}.  

In either case, the number of bifundamentals is $3M$, where $M$ is the number
of boxes.  Each bifundamental fermion has $R$-charge -1/3 in this case because
of the cubic superpotential, so the contribution to $c-a$ of the
$MN^2$ gaugino degrees of freedom cancels that of the $3MN^2$ matter fermions,
so $c-a=0$ in these elliptic brane box models.  It is not surprising that
the elliptic brane box models should satisfy the supergravity condition
since they have T-dual descriptions in terms of D3 branes at orbifold
singularities \cite{HSU,KS,HU,ibanez}.

If some of the boxes are made infinitely large in one direction, we obtain
the cylindrical brane box models described in \cite{HSU}.  Arguments similar
to those in the previous section show that these theories do not satisfy
the $c-a=0$ condition.

\section{Conclusions}
We have studied four dimensional ${\cal N}$=1 theories with quartic 
superpotentials and 
their brane description in Type IIA and Type IIB string theories.  
These theories can be obtained from ${\cal N}$=2 theories by integrating
out the adjoint chiral multiplet.  If the ${\cal N}$=2 theory is finite,
then the resulting ${\cal N}$=1 theory has marginal deformations along a
line of fixed points.  Type IIA elliptic models have Type IIB 
descriptions in terms of branes and orientifolds in singular
backgrounds.
We showed that a necessary condition for there to exist
a supergravity description of a theory, namely the equivalence of the Weyl 
and Euler
anomalies, is satisfied in the elliptic models except at special points along
the manifold of fixed points at which there are additional massless degrees
of freedom, in which case we could not reliably calculate the anomalies.  The 
condition $c-a=0$ also imposes severe restrictions
on the types of non-elliptic models that can have supergravity descriptions:
Any non-elliptic theory which can be obtained from 
Type IIA brane configurations with vanishing net sixbrane charge and NS5
brane linking numbers does not satisfy the supergravity condition
$c-a=0$.

\begin{center}{\bf Acknowledgments}\end{center}
We are happy to thank Bo Feng, Martin Gremm, 
Ken Intriligator, Angel Uranga and Cumrun Vafa for useful discussions.  This 
research is supported in 
part by the U.S. Department of Energy under cooperative agreement
\#DE-FC02-94ER40818.

\end{document}